%% file: main.tex
\documentclass[12pt]{article}
\usepackage[dvips]{graphicx}

\usepackage{amssymb}
\usepackage{amsmath}
\usepackage{latexsym}

\newcommand{\ccqqqqnn}{$\chi^{+} \chi^{-} \to qqqq \chi^{0} \chi^{0}$}
\newcommand{\misspt}{$^{\mathrm{miss}}p_{\mathrm{T}}$}

\newcommand{\wwqqqq}{$W^{+}W^{-} \to qqqq$}
\newcommand{\nnwwnnqqqq}{$\nu\bar{\nu}W^{+}W^{-} \to \nu\bar{\nu} qqqq$}
\newcommand{\eewweeqqqq}{$e^{+}e^{-} W^{+}W^{-} \to e^{+}e^{-} qqqq$}

\newcommand{\enwzenqqqq}{$e \nu_{e} W Z \to e \nu_{e} qqqq$}

\begin{document}

\begin{titlepage}
\begin{center}

\hfill KEK Preprint 2010-54 \\
\hfill IPMU11-0012 \\
\hfill TU-875 \\
\hfill UT-11-18 \\

\vspace{0.4cm}

{\large\bf Discrimination of New Physics Models \\
with the International Linear Collider}

\vspace{0.7cm}

{\bf Masaki Asano}$^{(a)}$\footnote{masano@hep-th.phys.s.u-tokyo.ac.jp},
{\bf Tomoyuki Saito}$^{(b)}$\footnote{saito@epx.phys.tohoku.ac.jp},
{\bf Taikan Suehara}$^{(c)}$\footnote{suehara@icepp.s.u-tokyo.ac.jp},\\
{\bf Keisuke Fujii}$^{(d)}$, {\bf R. S. Hundi}$^{(e)}$, {\bf Hideo Itoh}$^{(f)}$, 
{\bf Shigeki Matsumoto}$^{(g)}$, {\bf Nobuchika Okada}$^{(h)}$, 
{\bf Yosuke Takubo}$^{(d)}$, and {\bf Hitoshi Yamamoto}$^{(b)}$

\vspace{0.5cm}

{\it
$^{(a)}${Department of Physics, The University of Tokyo, Tokyo 113-0033, Japan} \\
$^{(b)}${Department of Physics, Tohoku University, Sendai 980-8578, Japan} \\
$^{(c)}${ICEPP, The University of Tokyo, Tokyo 113-0033, Japan} \\
$^{(d)}${High Energy Accelerator Research Organization (KEK), Tsukuba 305-0801, Japan} \\
$^{(e)}${
Department of Theoretical Physics, \\
Indian Association for the Cultivation of Science, Kolkata 700032, India} \\
$^{(f)}${ICRR, The University of Tokyo, Kashiwa 277-8582, Japan} \\
$^{(g)}${IPMU, TODIAS, The University of Tokyo, Kashiwa 277-8582, Japan } \\
$^{(h)}${Department of Physics and Astronomy, 
University of Alabama, \\ Tuscaloosa, AL 35487, USA} \\
}

\newpage
\vspace{0.7cm}

\abstract{The large hadron collider (LHC) is anticipated to provide signals of new physics at the TeV scale, which are likely to involve production of a WIMP dark matter candidate. The international linear collider (ILC) is to sort out these signals and lead us to some viable model of the new physics at the TeV scale. In this article, we discuss how the ILC can discriminate new physics models, taking the following three examples: the inert Higgs model, the supersymmetric model, and the littlest Higgs model with T-parity. These models predict dark matter particles with different spins, 0, 1/2, and 1, respectively, and hence comprise representative scenarios. Specifically, we focus on the pair production process, $e^+ e^- \to \chi^+ \chi^- \to \chi^0 \chi^0 W^+ W^-$, where $\chi^0$ and $\chi^\pm$ are the WIMP dark matter and a new charged particle predicted in each of these models. We then evaluate how accurately the properties of these new particles can be determined at the ILC and demonstrate that the ILC is capable of identifying the spin of the new charged particle and discriminating these models.}

\end{center}
\end{titlepage}

\setcounter{footnote}{0}

\input Introduction

\input Model

\input Framework

\input Analysis500

\input Analysis1000

\input Summary

\vspace{1.0cm}
\hspace{0.2cm} {\bf Acknowledgments}
\vspace{0.5cm}

The authors would like to thank all the members of the ILC physics subgroup \cite{Ref:subgroup} for useful discussions. This work is supported in part by the Creative Scientific Research Grant (No. 18GS0202) of the Japan Society for Promotion of Science (JSPS), JSPS Grant-in-Aid for Scientific Research (No. 22244031) and the JSPS Core University Program. This work is also supported in part by the Grant-in-Aid for Scientific research from the Ministry of Education, Science, Sports, and Culture (MEXT), Japan, No.~22244021 (M.~A.), the Grant-in-Aid for the Global COE Program Weaving Science Web beyond Particle-matter Hierarchy from the MEXT, Japan (M.~A.) and by the DOE Grants, No.~DE-FG02-10ER41714 (N.~O).


\end{document}

%% file: Introduction.tex
\section{Introduction} 
\label{sec:intro}

The standard model (SM) has no symmetry to protect the smallness of the scale of the electroweak symmetry breaking, hence, the Higgs mass receives quadratically divergent corrections, leading to the hierarchy problem. As a remedy for this problem, new physics beyond the SM is expected to appear at the TeV scale. 

The SM has, however, yet another problem. We know that about 23$\%$ of the energy density of the present universe is made up of unknown dark matter~\cite{Komatsu:2010fb} and that it played an important role in the formation of the large scale structure of the universe~\cite{Primack:2002th}. There is, however, no candidate for the dark matter in the SM. 

It seems plausible that the problem of the dark matter is also solved in the framework of the physics beyond the SM which solves the hierarchy problem. In the TeV scale physics, there are new particles which change the behavior of the quantum correction to the Higgs mass term. Some of the new particles would have the SU(2) charge  of the SM because it is related with the origin of the electroweak symmetry breaking. To solve the hierarchy problem without a fine-tuning, these have masses of $\mathcal{O}(100)$ GeV.
When the lightest of them is neutral and stable (e.g. the lightest neutralino in supersymmetric models with conserved R-parity), it is nothing but a Weakly Interacting Massive Particle (WIMP)~\cite{Jungman:1995df}. The WIMP is well known to be a good candidate for the dark matter, which naturally realizes the correct dark matter abundance in the present universe. Because of these attractive features, many new physics models at the TeV scale involving the WIMP dark matter candidate have been proposed.

One of the most important questions here is how to single out the new physics model at the TeV scale that consistently describes the results from energy frontier colliders such as the Large Hadron Collider (LHC) and the International Linear Collider (ILC). Uncovering the nature of the WIMP dark matter experimentally is of particular importance not only for particle physics but also for astrophysics and cosmology. The LHC experiments are now in operation where new physics signals are anticipated, which will guide us to narrow down possible models at the TeV scale. Being a hadron collider, the LHC is suitable to study colored new particles. It is, however, not an ideal place to do precision measurements of the properties of weakly interacting particles (non-colored particles) including the dark matter, while the ILC, being a lepton collider, has a great advantage for this purpose.

In this article, we investigate the possibility to discriminate new physics models at the ILC using the following process including the WIMP dark matter: $e^+e^- \rightarrow \chi^+ \chi^- \rightarrow \chi^0 \chi^0 W^+ W^-$, where $\chi^0$ and $\chi^{\pm}$ stand for the WIMP dark matter and a charged new particle predicted in each model. In our previous work~\cite{Asakawa:2009qb}, we have investigated the process in the framework of the littlest Higgs model with T-parity, and evaluated how accurately we can measure the properties of the new particles. We have shown there that the masses of $\chi^{\pm}$ and $\chi^0 $ can be determined to an accuracy of $1\%$ or better by locating the both endpoints of the energy distribution of the reconstructed $W$ bosons. It is also possible to determine the spin of $\chi^{\pm}$ and the structure of the interaction vertex between $\chi^\pm$, $\chi^0$, and $W$, through the observations of the angular distribution of $\chi^{\pm}$ and the polarization of $W$. The gauge charge of $\chi^{\pm}$ can also be measured, making use of polarized electron beam. Interestingly, the same process exists in various other new physics models at the TeV scale, and it turns out to be an extremely useful process to extract information on the new physics.

As the first step of our study to evaluate the ILC's potential to single out a viable new physics model, we investigate the possibilities to discriminate the following three models: the inert Higgs doublet model~\cite{Barbieri:2006dq}, the supersymmetric model~\cite{BookDrees}, and the littlest Higgs model with T-parity~\cite{ArkaniHamed:2002qy}. These models contain a WIMP dark matter particle with spin 0, 1/2, and 1, respectively~\cite{pastworks}. The masses of $\chi^\pm$ and $\chi^0$ are adjusted to coincide among different models. Although these models predict different cross sections for the $\chi^\pm$ pair production, we also force the cross sections to be a common value. We thus concentrate on the information related to the spin of the new charged particle for the discrimination of the new physics models.

This article is organized as follows. In the next section, we briefly review the new physics models used in our simulation study. Simulation framework such as representative points and simulation tools is presented in section 3. Details of the analysis to discriminate the new physics models are given in section 4, where the expected measurement accuracies of $\chi^\pm$ and $\chi^0$ properties are shown for each representative point. Section \ref{sec:summary} is devoted to summary.

%% file: Model.tex
\section{New Physics Models}
\label{sec:Model}

As already mentioned in the previous section, we concentrate on the process,
\begin{eqnarray}
e^+ e^- \to \chi^+ \chi^-  \to \chi^0 \chi^0 W^+ W^-,
\label{eq:XXW}
\end{eqnarray}
where the WIMP dark matter is denoted by $\chi^0$, while the new charged particle is $\chi^\pm$, and both particles are assumed to have odd charge under the Z$_2$ symmetry guaranteeing the stability of the dark matter. SM particles are assumed to have even charge under the symmetry. The interaction vertex between $\chi^0$, $\chi^\pm$, and $W^\mp$ exists in the most of new physics models at the TeV scale. On the other hand, the spins of $\chi^0$ and $\chi^\pm$ are dependent on the model. All possible combinations for the spins up to spin 1 are shown in Table~\ref{tab:spin_combi}.

\begin{table*}[t]
\center{
\begin{tabular}{c|c|c}
Particles & Spins & Representative Model \\
\hline
($\chi^{\pm}_S$, $\chi^0_S$) & (0, 0)     & Inert Higgs model \\
($\chi^{\pm}_F$, $\chi^0_F$) & (1/2, 1/2) & Supersymmetric model \\
($\chi^{\pm}_V$, $\chi^0_V$) & (1, 1)     & Littlest Higgs model \\
($\chi^{\pm}_V$, $\chi^0_S$) & (1, 0)     & No well-known models \\
($\chi^{\pm}_S$, $\chi^0_V$) & (0, 1)     & No well-known models \\
\hline
\end{tabular}
\caption{\small Spins of new particles $\chi^0$ and $\chi^\pm$ in various new physics models.}
\label{tab:spin_combi}
}
\end{table*}

At the ILC, $\chi^+$ and $\chi^-$ are produced in pairs through s-channel exchanges of photon ($\gamma$) and $Z$ boson, and the produced $\chi^\pm$ decays to $\chi^0$ and $W^{\pm}$. In addition, if there is another new particle which has a lepton number such as the sneutrino in the supersymmetric model or the heavy neutrino in the littlest Higgs model with T-parity, the diagram in which the new particle is exchanged in the t-channel contributes to the $\chi^\pm$ pair production. In our analysis, we simply assume that such a particle is heavy enough and ignore its contribution.

In our simulation study, we consider the inert Higgs doublet model, the supersymmetric model, and the littlest Higgs model with T-parity as benchmark models in which the $\chi^{\pm}$ has spin 0, 1/2, and 1, respectively, and develop the strategy to discriminate these models at the ILC. The crucial difference from the (1,0) or (0,1) models in Table~\ref{tab:spin_combi} only appears in what relates to the $\chi^{\pm}\chi^{0}W^{\mp}$ vertex (e.g. the shape of the energy distribution of $W$ bosons), so that the strategy developed in this article can be applied to the models with (1,0) or (0,1) spin combinations.

In the rest of this section, we briefly introduce the models used in our simulation study, focusing on interactions relevant to our analysis.

\subsection{Inert Higgs doublet model}

The inert Higgs doublet model~\cite{Barbieri:2006dq} is one of the two-Higgs-doublet models with unbroken Z$_2$ symmetry. One of the Higgs doublets transforms as $\phi \leftrightarrow - \phi$ under the discrete symmetry, while the other doublet and SM particles transform as ${\rm SM} \leftrightarrow {\rm SM}$. Because of the existence of the terms which break the custodial symmetry in the Higgs potential, the mass of the lightest Higgs boson could be as large as 500 GeV without conflicting with precision electroweak measurements. The fine-tuning between the Higgs boson mass and its radiative corrections, therefore, becomes mild compared to the SM with a light Higgs boson. In the model, neutral and charged components of the Z$_2$ odd Higgs boson, which is called the inert Higgs boson, play the role of the WIMP ($\chi^0_S$) and the new charged particle ($\chi^{\pm}_S$), both of which are scalar particles.

We focus on production and decay vertices of the new charged particle, which originate from gauge interactions:
\begin{eqnarray}
{\mathcal L}
&=&
i \left[ g_Z (1/2 - s_W^2) Z^\mu + e A^\mu \right]
\left[
\left(\partial_\mu \chi^+_S \right) \chi^-_S
-
\left(\partial_\mu \chi^-_S \right) \chi^+_S
\right] 
\nonumber \\
&&
+(g/2)
\left[
-\left(\partial^\mu \chi^+_S \right) \chi^0_S W^-_\mu
+\left(\partial^\mu \chi^0_S \right) \chi^+_S W^-_\mu
+ h.c.
\right],
\label{eq:IDM-V}
\end{eqnarray}
where, $e = \sqrt{4\pi\alpha}$ with $\alpha$ being the fine structure constant, $g$ is the SU(2)$_L$ gauge coupling constant, and $g_Z = e/(s_W c_W)$. The symbols $s_W$ and $c_W$ stand for $\sin\theta_W$ and $\cos\theta_W$, respectively, with $\theta_W$ being the Weinberg angle.

\subsection{Supersymmetric model}

Supersymmetry (SUSY) is the symmetry that relates particles of one spin to other particles that differ by half a unit of spin~\cite{BookDrees}. A new particle called superpartner is hence introduced for each SM particle in the SUSY model. It is known that the chiral symmetry guarantees the smallness of fermion masses. Since SUSY relates fermions to bosons, not only the smallness of fermion masses but also that of scalar masses are guaranteed, and the hierarchy problem of the SM disappears. In the SUSY model, if the R-parity is conserved, the lightest superpartner (LSP) is a good candidate for dark matter. One of the most plausible candidates for the LSP is the neutralino ($\chi^0_F$) which is a linear combination of superpartners of U(1) and neutral SU(2)$_L$ gauge bosons and neutral Higgs bosons. On the other hand, a new charged particle ($\chi^\pm_F$) is also predicted, namely, the chargino, which is a linear combination of superpartners of the charged SU(2)$_L$ gauge boson and the charged Higgs boson. The SUSY model therefore predicts fermionic new particles $\chi^0_F$ and $\chi^\pm_F$.

In this model, interactions needed for our simulation study have the form:
\begin{eqnarray}
{\mathcal L}
=
-g_Z \overline{\chi^-_F} \gamma^\mu
\left( N_L P_L + N_R P_R \right) \chi^-_F Z_\mu
-g \overline{\chi^-_F} \gamma^\mu
\left( C_L P_L + C_R P_R \right) \chi^0_F W^-_\mu
+ h.c.,
\label{eq:SUSY-V}
\end{eqnarray}
where, $P_L$ and $P_R$ are chirality projection operators. Coefficients $N_L$, $N_R$, $C_L$, and $C_R$ in front of the operators are determined by the mass matrices of neutralinos and charginos~\cite{BookDrees}, which depend on the details of the scenario. The values of the coefficients adopted in our simulation study are given in the next section.

\subsection{Littlest Higgs model with T-parity}

The littlest Higgs model with T-parity is based on a non-linear sigma model describing SU(5)/SO(5) symmetry breaking~\cite{ArkaniHamed:2002qy}, and the Higgs boson is regarded as one of the pseudo Nambu-Goldstone bosons arising from the breaking. The global symmetry SU(5) is not exact and is slightly broken due to the existence of explicit breaking terms, which are specially arranged to cancel quadratically divergent corrections to the Higgs mass term at 1-loop level. The quadratically divergent corrections appear, at most, at 2-loop level and the scale of the new physics can be as large as 10 TeV without the fine-tuning on the Higgs mass term~\cite{ArkaniHamed:2001nc}, thereby solving the
little hierarchy problem~\cite{Barbieri:1998uv}. Additionally, the implementation of the Z$_2$ symmetry called T-parity to the model has been proposed to evade severe constraints from electroweak precision measurements~\cite{Cheng:2003ju}. Due to the discrete symmetry, the lightest T-parity odd particle, which is the heavy photon ($\chi^0_V$), is a good candidate for dark matter. On the other hand, the charged new particle ($\chi^\pm_V$) which decays into $\chi^0_V$ and $W$ is the heavy $W$ boson. Both $\chi^0_V$ and $\chi^\pm_V$ are massive vector bosons, and acquire their masses through the SU(5)/SO(5) braking.

In this model, interactions relevant to our simulation study are given by
\begin{eqnarray}
 {\mathcal L} &=& i g
 \left[
  ~~~
  \left(c_W Z + s_W A\right)_\mu \chi_{V\nu}^+
  \left(\partial^\mu \chi_V^{-\nu} - \partial^\nu \chi_V^{-\mu}\right)
 \right.
 \nonumber \\
 &&\qquad
 -
 (c_W Z + s_W A)_\mu \chi_{V\nu}^-
 \left(\partial^\mu \chi_V^{+\nu} - \partial^\nu \chi_V^{+\mu}\right)
 \nonumber \\
 &&\qquad
 +
 \partial_\mu \left(c_W Z + s_W A\right)_\nu
 \left(\chi_V^{+\mu}\chi_V^{-\nu} - \chi_V^{-\mu} \chi_V^{+\nu}\right)
 \nonumber \\
 &&\qquad
 +
 s_H W^+_\mu \chi_{V\nu}^-
 \left(
  \partial^\mu \chi_V^{0 \nu} - \partial^\nu \chi_V^{0 \mu}
 \right)
 -
 s_H W^+_\mu \chi_{V \nu}^0
 \left(\partial^\mu \chi_V^{- \nu} - \partial^\nu \chi_V^{- \mu}\right)
 \nonumber \\
 &&\qquad
 +
 s_H \partial_\mu W^+_\nu
 \left(
  \chi_V^{0 \nu} \chi_V^{- \mu} 
  -
  \chi_V^{0 \mu} \chi_V^{- \nu}
 \right)
 -
 s_H W_\mu^- \chi^+_{V \nu}
 \left(
  \partial^\mu \chi_V^{0 \nu} 
  -
  \partial^\nu \chi_V^{0 \mu}
 \right)
 \nonumber \\
 &&\qquad
 +
 s_H W_\mu^- \chi_{V \nu}^0
 \left(\partial^\mu \chi_V^{+\nu} - \partial^\nu \chi_V^{+\mu}\right)
 \left.
  -
  s_H \partial_\mu W_\nu^-
  \left(
   \chi_V^{0 \nu} \chi_V^{+\mu}
   -
   \chi_V^{0 \mu} \chi_V^{+\nu}
  \right)
  ~~~
 \right],
 \label{eq:LHT-V}
\end{eqnarray}
where $s_H = \sin\theta_H$ with $\theta_H$ being the mixing angle between neutral heavy gauge bosons and determined by the mass matrix of the bosons. The value of $\theta_H$ used in our simulation study is also given in the next section.

%% file: Framework.tex
\section{Simulation framework}
\label{sec:framework}

In this section, we summarize the simulation framework such as representative points used in our analysis, strategy to discriminate the new physics models discussed in the previous section, and tools used in the simulation study.

\subsection{Representative points}

Mass spectrum of the WIMP dark matter ($\chi^0$) and the new charged particle ($\chi^\pm$) used in our analysis is shown in Table~\ref{tab:sample points}. This mass spectrum is adopted in all the new physics models. Though the three new physics models predict different cross section values for  $\chi^\pm$ pair production, we use a common value for the cross section with 100\% branching ratio for the decay $\chi^\pm \to \chi^0 W^\pm$. Two cross section values are considered, 40 and 200 fb, as shown in Table~\ref{tab:sample points}. We therefore call the models the inert Higgs-like (IH-like), supersymmetric-like (SUSY-like), and littlest Higgs-like (LHT-like) models, respectively, in the following discussions.

\begin{table*}[t]
\center{
\begin{tabular}{c|cccc}
& $m_{\chi^\pm}$ [GeV] & $m_{\chi^0}$ [GeV] & Cross section [fb] & $\sqrt{s}$ [GeV] \\
\hline
Point I  & 232 & 44.0 & 40 \& 200 &  500 \\
Point II & 368 & 81.9 & 40 \& 200 & 1000 \\
\hline
\end{tabular}
\caption{\small Representative points used in our simulation study.}
\label{tab:sample points}
}
\end{table*}

In the inert Higgs model, the structures of interaction vertices $\chi^+$-$\chi^-$-$Z(\gamma)$ and $\chi^\pm$-$\chi^0$-$W^\mp$ are completely fixed\footnote{The cross section for the $\chi^\pm$ production in the inert Higgs model is 3.51 (6.85) fb at Point I (II).  These cross sections are much smaller than the corresponding cross sections in the supersymmetric model and the littlest Higgs model with T-parity (see footnotes 2 and 3). This is because the production cross section is proportional to $\beta^3$, in addition to the fact that the production rates for scalar particles are usually smaller than those for spin 1/2 fermions or vector bosons.}. On the other hand, in the supersymmetric model, there are parameters to be fixed for the vertices: $N_{L(R)}$ and $C_{L(R)}$. With the SUSY parameters, $m_0 =$ 5 (10) TeV, $M_1 =$ 44.5 (81.0) GeV, $M_2 =$ 234 (369) GeV, $\mu =$ 1 (1) TeV, and $\tan\beta =$ 10 (10) at the TeV scale, the masses of the lightest neutralino ($\chi^0$) and chargino ($\chi^\pm$) turn out to be 44.0 (81.9) GeV and 232 (368) GeV, respectively, at Point I (II). Using the parameters, the ratio of the coefficients between $N_L$ and $N_R$ is determined to be $N_L/N_R =$ 0.992 (1.00), while $C_L/C_R$ is 1.36 (1.31). We adopt these coefficients in the SUSY-like model\footnote{Cross section for the $\chi^\pm$ production in the supersymmetric model is 414 (201) fb at Point I (II).}. As in the supersymmetric model, there is a parameter to be fixed for the vertices in the littlest Higgs model with T-parity: $\theta_H$. By choosing the vacuum expectation value of the SU(5)/SO(5) symmetry breaking to be 375 (580) GeV, we can adjust the masses of $\chi^0$ and $\chi^\pm$ to be 44.0 (81.9) GeV and 232 (368) GeV. With this vacuum expectation value, the angle $\theta_H$ is determined as $\tan\theta_H =$ $-0.0525$ ($-0.0246$), and we use these values in the LHT-like model\footnote{Cross section for the $\chi^\pm$ production in the littlest Higgs model with T-parity is 364 (693) fb at Point I (II).}.

\subsection{Simulation strategy}

Since the dark matter will escape without detection, the measurement of the new physics models at the TeV scale (IH-like, SUSY-like, and LHT-like models) is not straightforward. In the paper, in order to discriminate the new physics models, we focus on the following three physical quantities, (i) the energy distribution of the $W$ boson, (ii) the angular distribution of the new charged particle $\chi^\pm$, and (iii) the threshold behavior of the cross section for the $\chi^\pm$ pair production. These quantities are relevant to kinematics of the process and spin information of the new charged particles. In this subsection, we discuss how measurements of these quantities work for discrimination of the new physics models.

\subsubsection{Energy distribution of $W$}

Solving the kinematics of the new physics process $e^+ e^- \rightarrow \chi^+ \chi^- \rightarrow \chi^0 \chi^0 W^+ W^-$, we find the maximum and the minimum of the $W$ energy ($E_{\rm max}$ and $E_{\rm min}$) given by
\begin{eqnarray}
&&
E_{\rm max}
=
\gamma_{\chi^\pm} E_W^* + \beta_{\chi^\pm} \gamma_{\chi^\pm} p^*_W,
\nonumber \\
&&
E_{\rm min}
=
\gamma_{\chi^\pm} E_W^* - \beta_{\chi^\pm} \gamma_{\chi^\pm} p^*_W,
\end{eqnarray}
where $\beta_{\chi^\pm}$ ($\gamma_{\chi^\pm}$) is the $\beta$ ($\gamma$) factor of $\chi^\pm$ in the laboratory frame, while $E^*_W$ ($p^*_W$) is the energy (momentum) of the $W$ boson in the rest frame of $\chi^\pm$. The energy $E^*_W$ is given as $(M_{\chi^\pm}^2 + M_W^2 - M_{\chi^0}^2)/(2M_{\chi^0})$. As a result, both masses of $\chi^\pm$ and $\chi^0$ can be estimated from the edges of the distribution of the reconstructed $W$ boson energy.

\subsubsection{Angular distribution of the $\chi^\pm$ production}

The production angle of the new charged particle $\chi^\pm$ can be reconstructed up to two-fold ambiguity from the reconstructed $W$ boson momenta. The distribution of the reconstructed $\chi^\pm$ production angle allows us to investigate the property of $\chi^\pm$, because it depends on the spin of $\chi^\pm$. The angular distribution in each case of the new physics models (IH-like, SUSY-like, or LHT-like model) turns out to be
\begin{eqnarray}
\frac{d \sigma}{d (\cos \theta)}
\propto
\left\{
\begin{array}{ll}
1 - \cos^2 \theta & ({\rm for~IH-like}), \\
(1 + x/4) - (1 - x/4) \cos^2 \theta & ({\rm for~SUSY-like}), \\
(1 + x + x^2/12) - (1  - x/3 + x^2/12) \cos^2 \theta & ({\rm for~LHT-like}),
\end{array}
\right.
\label{Angular distribution}
\end{eqnarray}
where $x = s/M_{\chi^{\pm}}^2$ with $s$ being the center of mass energy and $\theta$ is the angle between the $\chi^\pm$ momentum and the beam axis. As demonstrated in the following sections, the angular distribution turns out to be a powerful tool to discriminate the new physics models.

\subsubsection{Threshold behavior of the $\chi^\pm$ production}

Since the $\chi^\pm$ pair production occurs in energetic $e^+ e^-$ collision through $s$-channel gauge boson exchanges, the total angular momentum along the beam axis in the initial state is one. The orbital angular momentum, therefore, has to be one (P-wave) when $\chi^\pm$ is a scalar particle, which leads to the behavior of the cross section $\sigma \propto (s - 4 M_{\chi^\pm}^2)^{3/2}$ in the threshold region $s \sim 4 M_{\chi^\pm}^2$. On the other hand, when the $\chi^\pm$ is a Dirac fermion, it can be produced with the S-wave, leading to the threshold behavior $\sigma \propto (s - 4 M_{\chi^\pm}^2)^{1/2}$. In the case of the vector $\chi^\pm$, the situation is more complicated. Since the $\chi^\pm$ in the littlest Higgs model is a gauge boson, the production vertex is coming from gauge self-interactions. In addition, there is also a vertex between the SM gauge bosons and would-be Nambu-Goldstone bosons absorbed in the longitudinal mode of $\chi^\pm$. In both cases, the final state with the total spin 1 cannot be composed by the vertices alone, which leads to the threshold behavior $\sigma \propto (s - 4 M_{\chi^\pm}^2)^{3/2}$ in the vector $\chi^\pm$ production. The threshold behavior of the $\chi^+$ production can therefore be used to discriminate the SUSY-like model from the rest.

\subsection{Simulation tools}

\subsubsection{Event generation}

For both Points I \& II, we generated signal events by using the Physsim~\cite{physsim} package. In this package, helicity amplitudes are calculated using the HELAS library~\cite{helas}, which deals with the effect of gauge boson polarizations properly. Phase space integration and the generation of parton four-momenta are performed by BASES/SPRING~\cite{bases}. Parton showering and hadronization are carried out by using PYTHIA6.4~\cite{pythia}, where final-state tau leptons are decayed by TAUOLA ~\cite{tauola} in order to handle their polarizations correctly.

For Point I, we generated SM background events by using the matrix element generator WHIZARD~\cite{whizard} 1.40 with PYTHIA 6.205 for the hadronization. It is the standard sample for Letter of Intent (LoI)~\cite{:2010zzd, Aihara:2009ad} study of ILC detector concepts, covering the whole SM processes with 12 million events in total. In contrast, for Point II, SM background events were generated using the Physsim package.

In all the generated samples, initial-state radiation and beamstrahlung effects are included. We ignore the finite crossing angle between the electron and positron beams and assume no initial beam polarizations\footnote{In general signal and background cross sections depend on beam polarization combination. In this study, however, we use no beam polarization so as to keep our study as model-independent as possible. If some enhancement is observed for a certain beam polarization combination, we can certainly use it to increase our signal statistics. In the case of the maximum enhancement, where the signal process is through a single $e^-$ and $e^+$ polarization combination, the enhancement is a factor of 2.26 for the nominal beam polarizations of 80\% in electrons and 30\% in positrons. Since most of the background processes are enhanced with the left-handed electrons, the background can be significantly suppressed if the signal process favors the right-handed electrons.}.

\subsubsection{Detector simulation}

For Point I, a full simulation code~\cite{mokka,marlinreco}, developed for the International Large Detector (ILD)~\cite{:2010zzd}, is used  for the Monte-Carlo (MC) simulation and event reconstruction. The standard geometry for the ILD LoI study is used for the detector simulation. The geometry includes a time projection chamber with silicon devices for tracking and vertexing, and highly granular electromagnetic and hadronic calorimeters for particle flow calorimetry along with a 3.5 Tesla magnetic field. The central part of the reconstruction is a particle flow algorithm~\cite{pandora}, which reconstructs individual charged and neutral particles from tracks and calorimeter clusters. The reconstructed particles are clustered into 4-jet configuration using the Durham algorithm~\cite{durham}. A neural-net based flavor tagging algorithm~\cite{lcfi} is applied to the jets after the jet clustering.

\begin{table}
 \center{
  \begin{tabular}{lcr}
   \hline
   Detector & Performance & Coverage \\
   \hline
   Vertex detector &
   $\delta_{b} \leq 5 \oplus 10/ p \beta \sin^{3/2}\theta$ ($\mu$m) &
   $|\cos\theta| \leq 0.93$
   \\
   Central drift chamber &
   $\delta p_{t}/p_{t}^{2} \leq 5 \times 10^{-5}$ (GeV/c)$^{-1}$ &
   $|\cos\theta| \leq 0.98$
   \\
   EM calorimeter &
   $\sigma_{E}/E = 17\% / \sqrt{E} \oplus 1\%$ &
   $|\cos\theta| \leq 0.99$
   \\
   Hadron calorimeter &
   $\sigma_{E}/E = 45\% / \sqrt{E} \oplus 2\%$ &
   $|\cos\theta| \leq 0.99$
   \\
   \hline
  \end{tabular}
 }
 \caption{\small Detector parameters used in the Point II study.}
 \label{tb:GLD}
\end{table}

For Point II, we use a fast simulator code~\cite{QSim}, which implements the GLD geometry and other detector performance related parameters~\cite{glddod}. In the simulator, hits by charged particles at the vertex detector and track parameters at the central tracker are smeared according to their position resolutions, taking into account correlations due to off-diagonal elements in the error matrix. Since calorimeter signals are simulated in individual segments, a realistic simulation of cluster overlap is possible. Track-cluster matching is performed for the hit clusters in the calorimeter in order to achieve the best energy flow measurements. The resulting detector performance in our simulation study is summarized in Table \ref{tb:GLD}.

%% file: Analysis500.tex
\section{Results from simulation study} \label{sec:analysis}
In this section, we present results from our simulation study for $e^+e^- \to \chi^+ \chi^- \to \chi^0 \chi^0 W^+ W^- $ process in the case of the IH-like, the SUSY-like, and the LHT-like models. We take two cross section points: $\sigma_s =$ 200 fb and 40 fb as examples. The simulation was performed at $\sqrt{s} =$ 500 GeV for Point I, and 1 TeV for Point II. An integrated luminosity $\mathcal{L}_\mathrm{int} = 500$ fb$^{-1}$ is assumed in each point for all the following study except for threshold scans.

\subsection{Study for Point I with $\sqrt{s} = 500$ GeV full simulation}

\subsubsection{Signal Selection}

Point I employs $m_{\chi^\pm} = 232$ GeV and $m_{\chi^0} = 44.0$ GeV, which can be investigated at the $\sqrt{s} =$ 500 GeV ILC.
We select signal events with both $W$'s decaying into two quarks ($qqqq$ events), whose branching fraction is about 46\%,
since the $W$ energies must be fully reconstructed for the mass determination and the production angle reconstruction.
The target event topology is thus 4-jets with missing momentum.
All SM processes with up to 6 particles in the final state are used in the analysis as background.
The dominant contribution is
the $W$-pair production with fully hadronic decays,
the $WWZ$ processes with the $Z$ decaying to a neutrino pair,
the top-pair production with one $W$ decaying leptonically, and 
$\gamma\gamma \rightarrow WW$ processes.
The SM Higgs ($ZH, \nu\nu{}H$) processes with $m_H = 120$ GeV and 
semi-leptonic signal processes are also included.

To reject a major part of the SM and the semi-leptonic decay background,
we applied primary selection cuts to all samples as follows.
(i) The number of tracks should be larger than $20$ and
each jet has to contain at least two tracks in order to eliminate pure leptonic events.
(ii) The visible energy of the event, $E_{\mathrm{vis}}$, should be between $80$ and $400$ GeV,
which can remove most of 2-photon and 2, 4, and 6-quark events.
(iii) Each jet should have a reconstructed energy of at least 5~GeV and a polar angle $\theta$ 
  fulfilling $|\cos\theta| < 0.99$ to ensure proper jet reconstruction.
(iv) The distance parameter
  of the Durham jet algorithm~\cite{durham} for which the event changes from 4-jet to 3-jet
  configuration, $y_{34}$, should be larger than 0.001 in order to reject most of 2-jet events.
(v) No lepton candidate with an energy larger than 25~GeV is allowed
  in order to suppress semi-leptonic events.
(vi) $|\cos\theta|$ of the missing momentum should be smaller than 0.9
and $|\cos\theta|$ summed up for all jets should be smaller than $2.6$
in order to eliminate most of the SM events which are concentrated in the forward region.
(vii) The neural-net output of $b$-tag probability summed up for all jets should be smaller than 1
to remove events with $b$ quarks.

After the primary selection,
a constrained kinematic fit~\cite{kinfitnote}, which 
requires the two dijet masses of the event to be equal, was performed on each event. All three possible jet pairings are tested
and the pairing with the least $\chi^2$ value for the kinematic fit is selected for the following analysis.

Secondary selection cuts were applied after the kinematic fit as follows.
(viii) The kinematic fit constraining the two dijet masses to be equal should converge for at 
least one jet pairing to ensure integrity of the fit result.
(ix) The di-jet mass obtained by the kinematic fit should be between $65$ and $95$ GeV
to select two-$W$ events.

\begin{table}
\begin{center}
                \begin{tabular}{|c|l|r|r|} \hline
           & Process                                                & \# of events & \# of events after cuts \\ \hline
           & IH-like (hadronic decay)                               & 46,815    & 27,837    \\ 
Signal     & SUSY-like (hadronic decay)                             & 45,550    & 26,578    \\ 
           & LHT-like (hadronic decay)                              & 46,644    & 27,631    \\ \hline
           & IH-like (other decay)                                  & 53,186    &   122    \\ 
Model bkg. & SUSY-like (other decay)                                & 54,462    &   104    \\ 
           & LHT-like (other decay)                                 & 53,355    &   212    \\ \hline
           & $qqqq$ ($WW$, $ZZ$)                                    & $1.88\times 10^6$ &  3,218    \\ 
           & $qq\ell\nu$ ($WW$)                                     & $2.35\times 10^6$ &  1,883    \\ 
           & $qqqq\nu\nu$ ($WWZ$)                                   & 4,158     &   681    \\ 
SM bkg.    & $qqqq\ell\nu$ ($tt$)                                   & 125,205   &   626    \\ 
           & $\gamma\gamma\rightarrow{}qqqq$                        & 26,356    &   509    \\ 
           & $qq$                                                   & $6.29\times 10^6$ &   373    \\ 
           & SM Higgs (120 GeV)                                     & 56,967    &    61    \\ 
           & Other background                                       & $3.44\times 10^9$ &   338    \\ \hline
                \end{tabular}
\caption{Event numbers before and after the selection cuts, normalized to $\mathcal{L}_\mathrm{int} = 500$ fb$^{-1}$
and $\sigma_s = 200$ fb in the Point I study.}
\label{tbl:cutstat500}
\end{center}
\end{table}

\begin{figure}[p]
	\begin{center}
		\includegraphics[height=0.3\textheight]{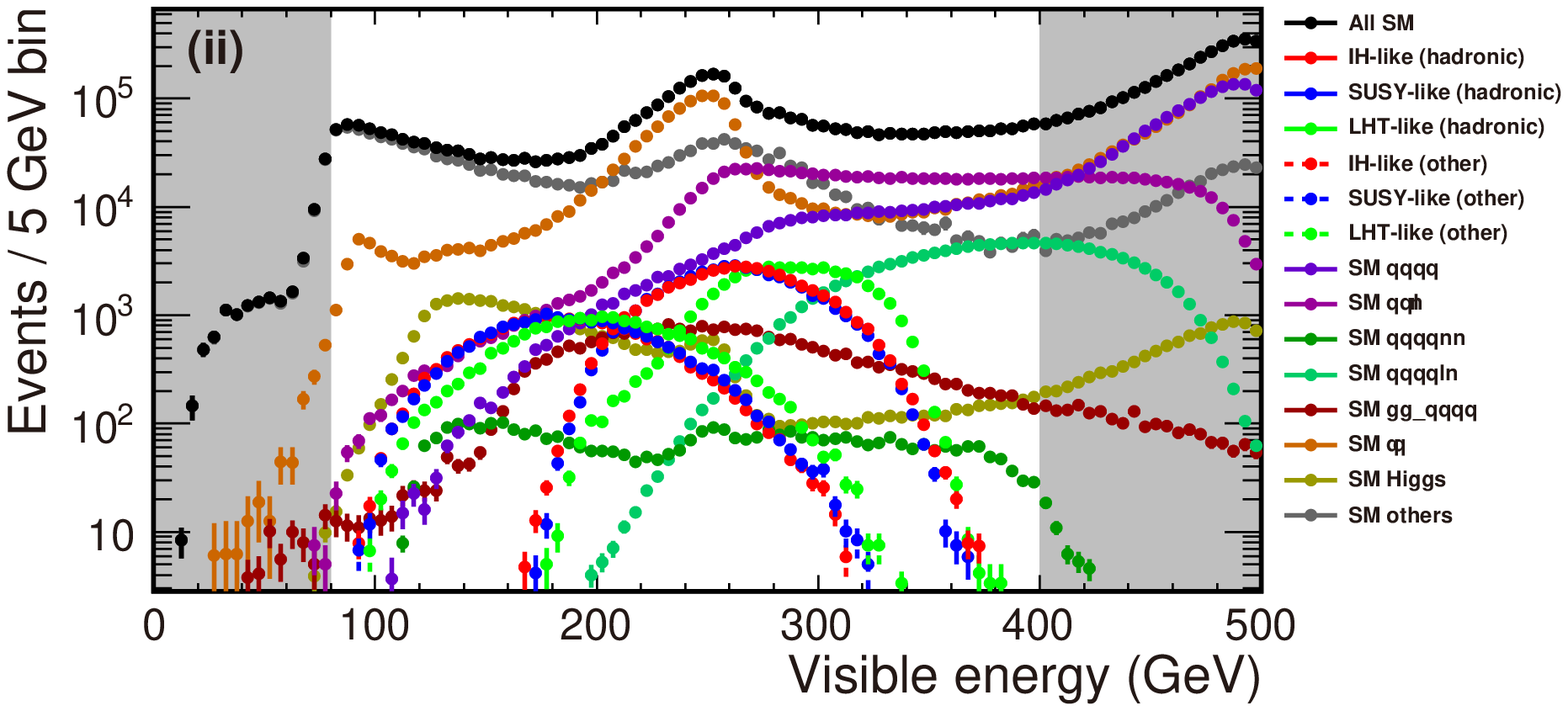}
		\includegraphics[height=0.3\textheight]{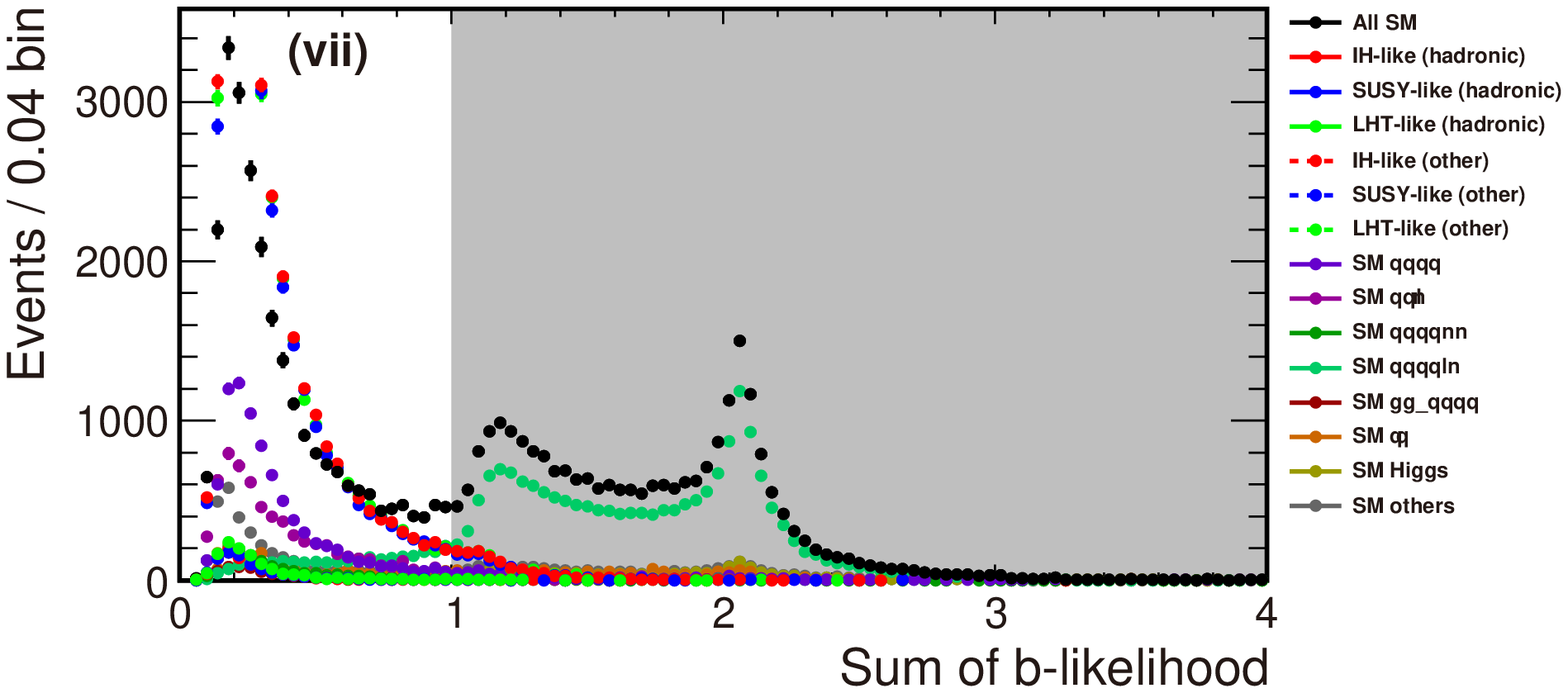}
		\includegraphics[height=0.3\textheight]{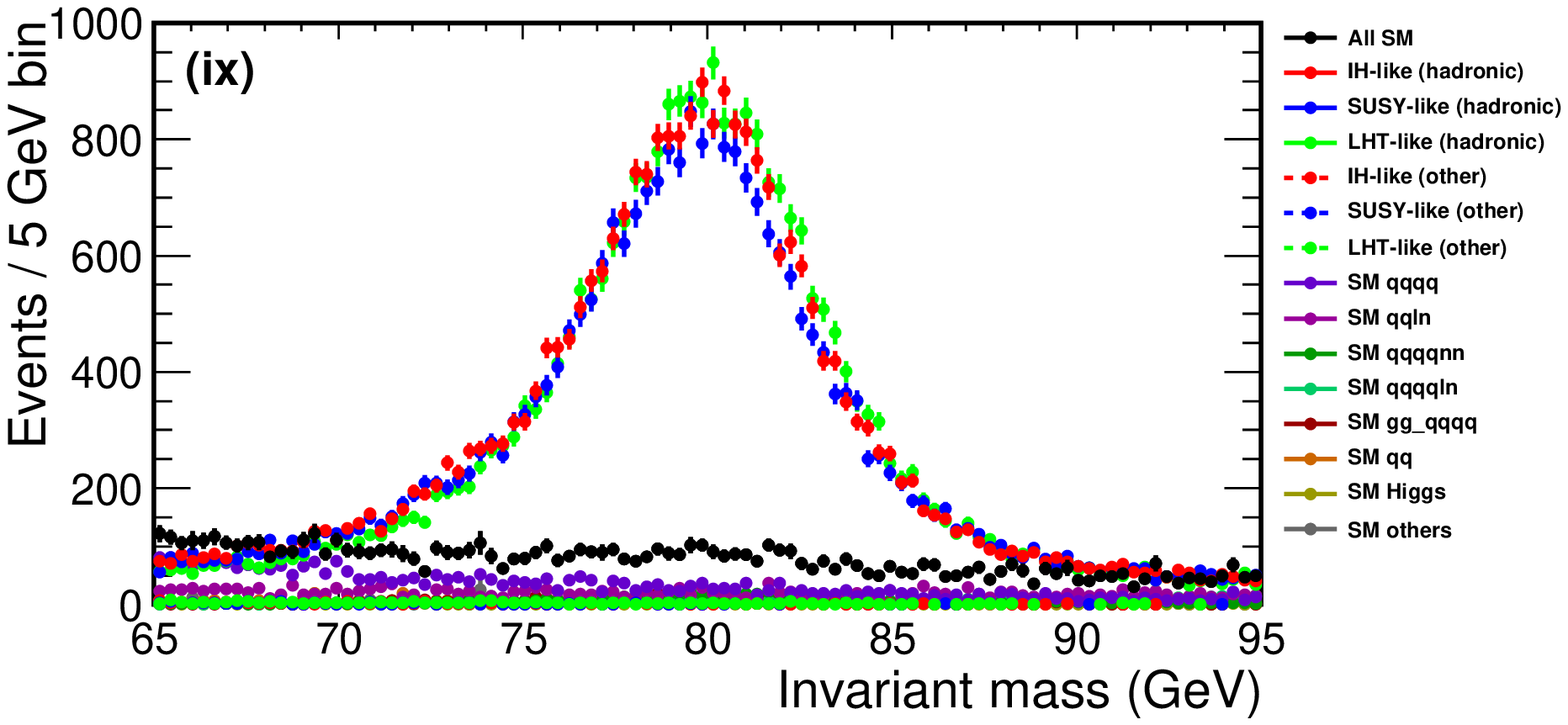}
	\end{center}
	\caption{Cut plots with $\sigma_s = 200$ fb, $\mathcal{L}_\mathrm{int} = 500$ fb$^{-1}$ in the Point I study. The labels (ii), (vii), (ix) correspond to
the cuts described in the text with the same labels. Grayed regions are cut out with the selection.}
	\label{fig:cutplot500}
\end{figure}

The effect of these cuts is summarized in Table~\ref{tbl:cutstat500} and
the distributions of some cut variables are shown in Figure \ref{fig:cutplot500}.
Clear peaks at di-jet masses of 80 GeV can be seen in the signal distributions of 
Figure \ref{fig:cutplot500} (ix), which are from two $W$ bosons.
Acceptances of signal events after the cuts are 59.5, 58.3, and 59.2 \% for the IH-like, the SUSY-like and the LHT-like
models, respectively. Signal purities after the cuts are 77.4, 76.6, and 77.3 \%
in the $\sigma_s = 200$ fb case and 40.7, 39.5 and 40.5 \% in the $\sigma_s = 40$ fb case, respectively.

\subsubsection{Mass Determination}

\begin{figure}[p]
	\begin{center}
		\includegraphics[height=0.3\textheight]{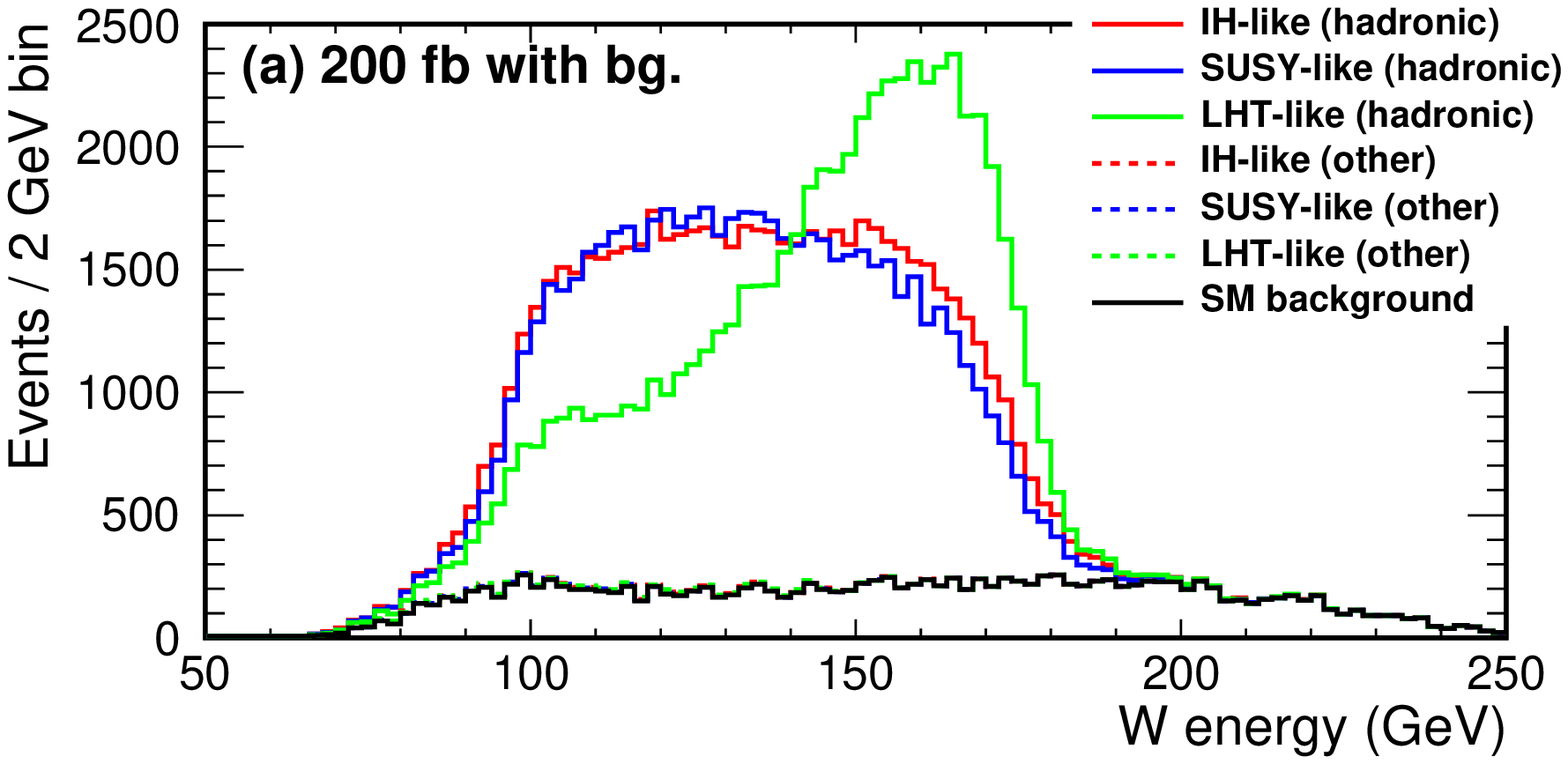}
		\includegraphics[height=0.3\textheight]{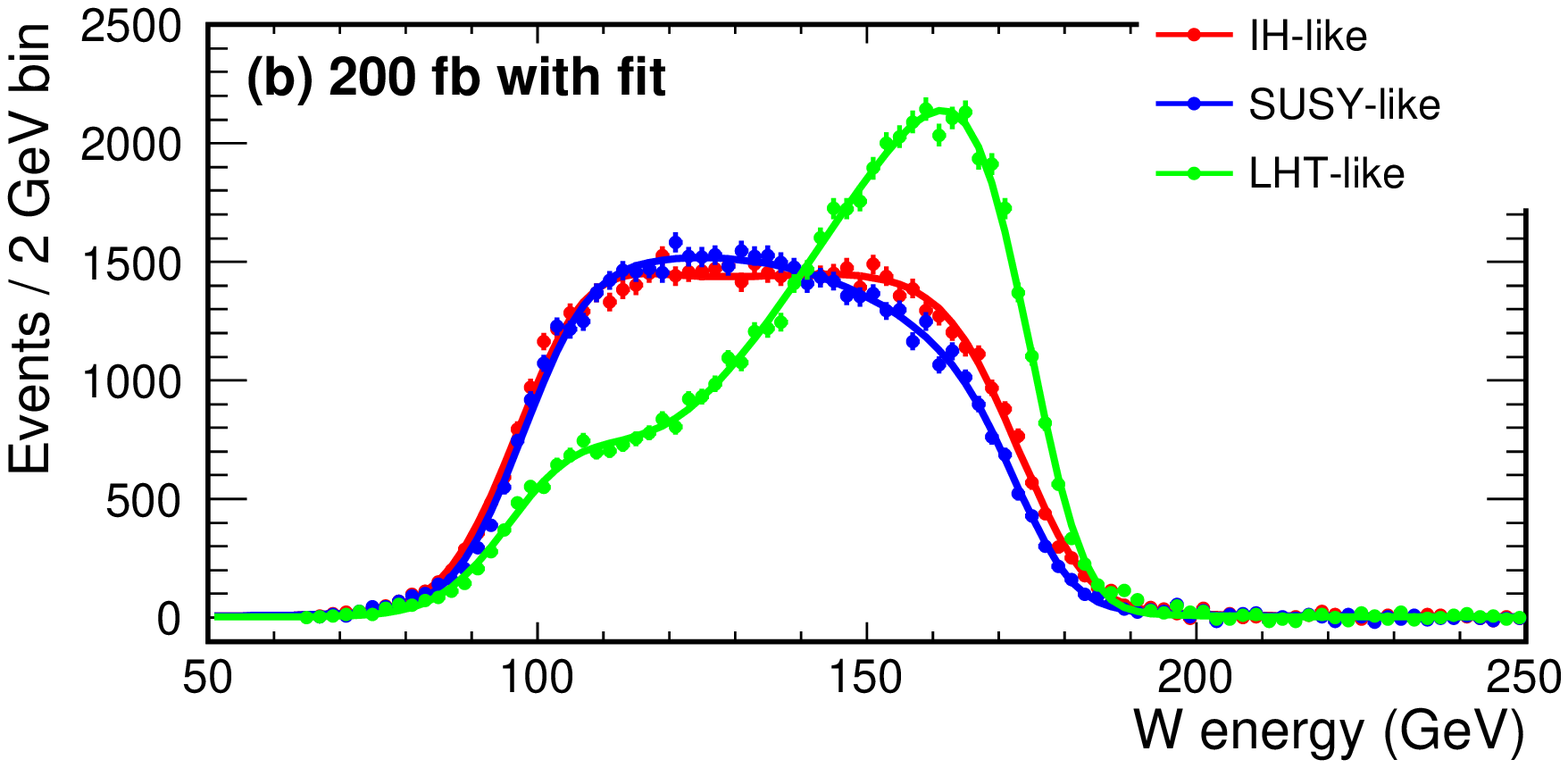}
		\includegraphics[height=0.3\textheight]{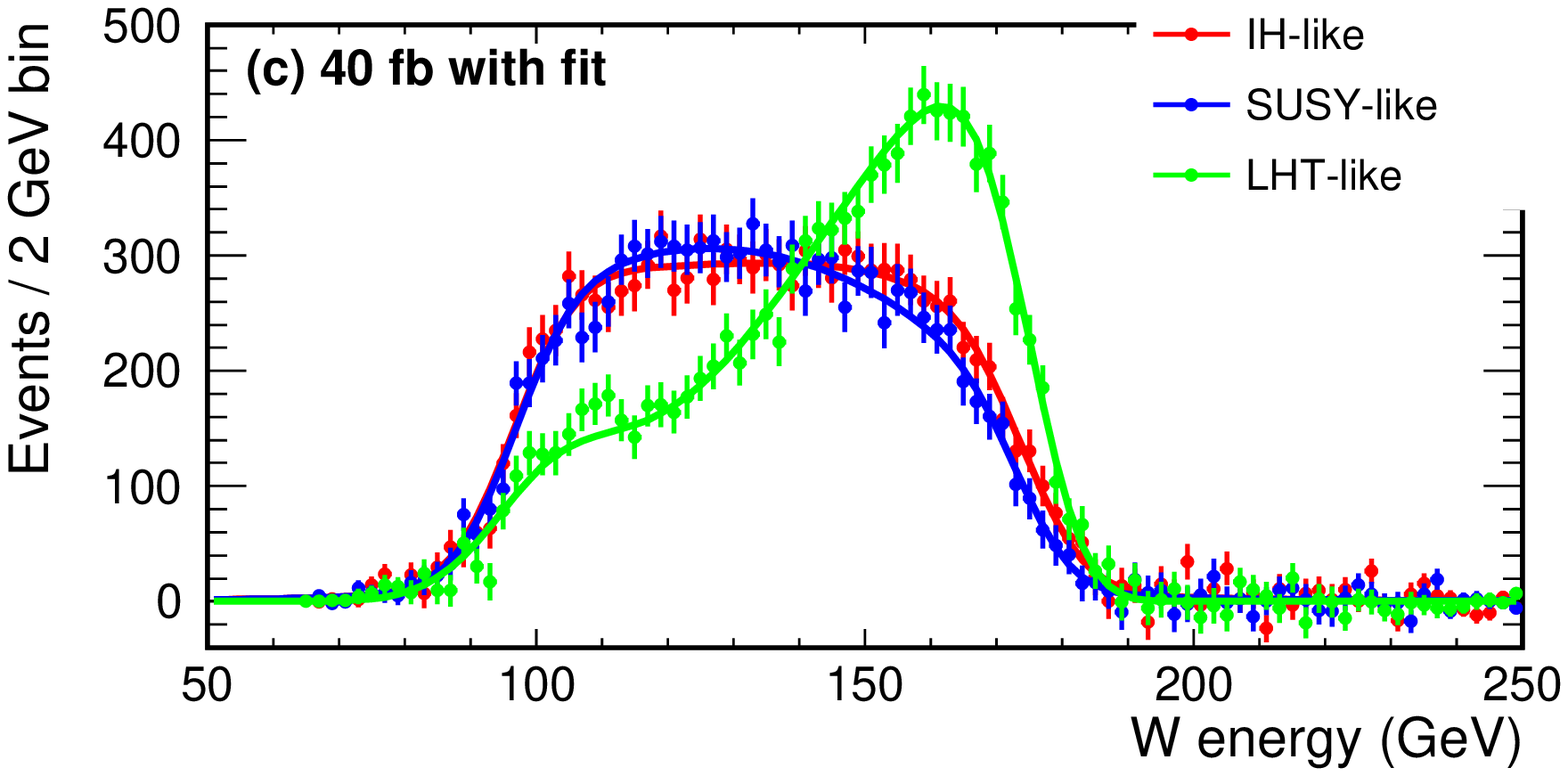}
	\caption{(a)$W$ energy distributions for signal ($\sigma_s = 200$ fb) and background with $\mathcal{L}_\mathrm{int} = 500$ fb$^{-1}$ in the Point I study.
		(b),(c) Results of the mass fit for $\sigma_s = 200$ fb and 40 fb after background subtraction, respectively.}	
	\label{fig:wmass500}
	\end{center}
\end{figure}

The masses of new particles can be obtained via the energy spectrum of the $W$ boson candidates.
The energy of the $W$ bosons has
upper and lower kinematic limits, from which the masses of the new particles can be derived.
Figure \ref{fig:wmass500}(a) shows the $W$ energy spectrum for each model on top of the SM background.
Clear edges can be seen in the distribution of every model.

The edge positions are obtained by a fit using an empirical function with kinematical edges.
The analysis was done by three steps as follows. (1) Determine shape paremeters of the fitting function
with a high-statistical sample (about 1 million events per model).
(2) Determine the edge positions and the normalization factor
(three free parameters) by a fit to a sample with the signal cross section (200 fb and 40 fb).
(3) Calculate the masses of $\chi^\pm$ and $\chi^0$ with the obtained edge positions.
The measurement of the edge positions are assumed to be statistically independent.

Figures \ref{fig:wmass500}(b) and \ref{fig:wmass500}(c) give the fitting results of $\sigma_s = 200$ fb and 40 fb with $\mathcal{L}_\mathrm{int} = 500$ fb$^{-1}$, respectively.
The fitting results are summarized in Table \ref{tbl:massfit500}.
While the central values of the fitting results deviate from the expected masses,
they can be corrected using Monte-Carlo samples in the real experiment.


\begin{table}
\begin{center}
\begin{tabular}{|l|r|r|r|}
\hline
 & Physics model & $\sigma_s = 200$ fb & $\sigma_s = 40$ fb \\
\hline
$M_{\chi^{\pm}}$ (GeV) & IH-like   & 232.9 $\pm$ 0.1 & 231.8 $\pm$ 0.4 \\
                       & SUSY-like & 232.7 $\pm$ 0.1 & 232.2 $\pm$ 0.5 \\
                       & LHT-like  & 232.1 $\pm$ 0.1 & 231.5 $\pm$ 0.5 \\ \hline 
$M_{\chi^{0}}$ (GeV)   & IH-like   &  44.2 $\pm$ 0.6 &  46.2 $\pm$ 1.9 \\
                       & SUSY-like &  43.6 $\pm$ 0.7 &  45.8 $\pm$ 2.3 \\
                       & LHT-like  &  43.8 $\pm$ 0.5 &  45.9 $\pm$ 1.8 \\ 
\hline
\end{tabular}
\end{center}
\caption{Measurement accuracies for the masses of $\chi^{\pm}$ and $\chi^{0}$ with $\mathcal{L}_\mathrm{int} = 500$ fb$^{-1}$ in the Point I study.}
 \label{tbl:massfit500}
\end{table}

In the $W$ energy distributions, we can see a clear difference among three models,
which may be used for the model separation.
However, the difference is considered to be coming from the vertex structures of interactions of the specific models
and not from the general spin structure,
thus we do not use this difference so as to keep this study model-independent.

\subsubsection{Angular Distribution for $\chi^\pm$ Pair Production}

\begin{figure}[p]
	\begin{center}
	\begin{minipage}[t]{.48\textwidth}
	\begin{center}
    \includegraphics[width=.8\textwidth]{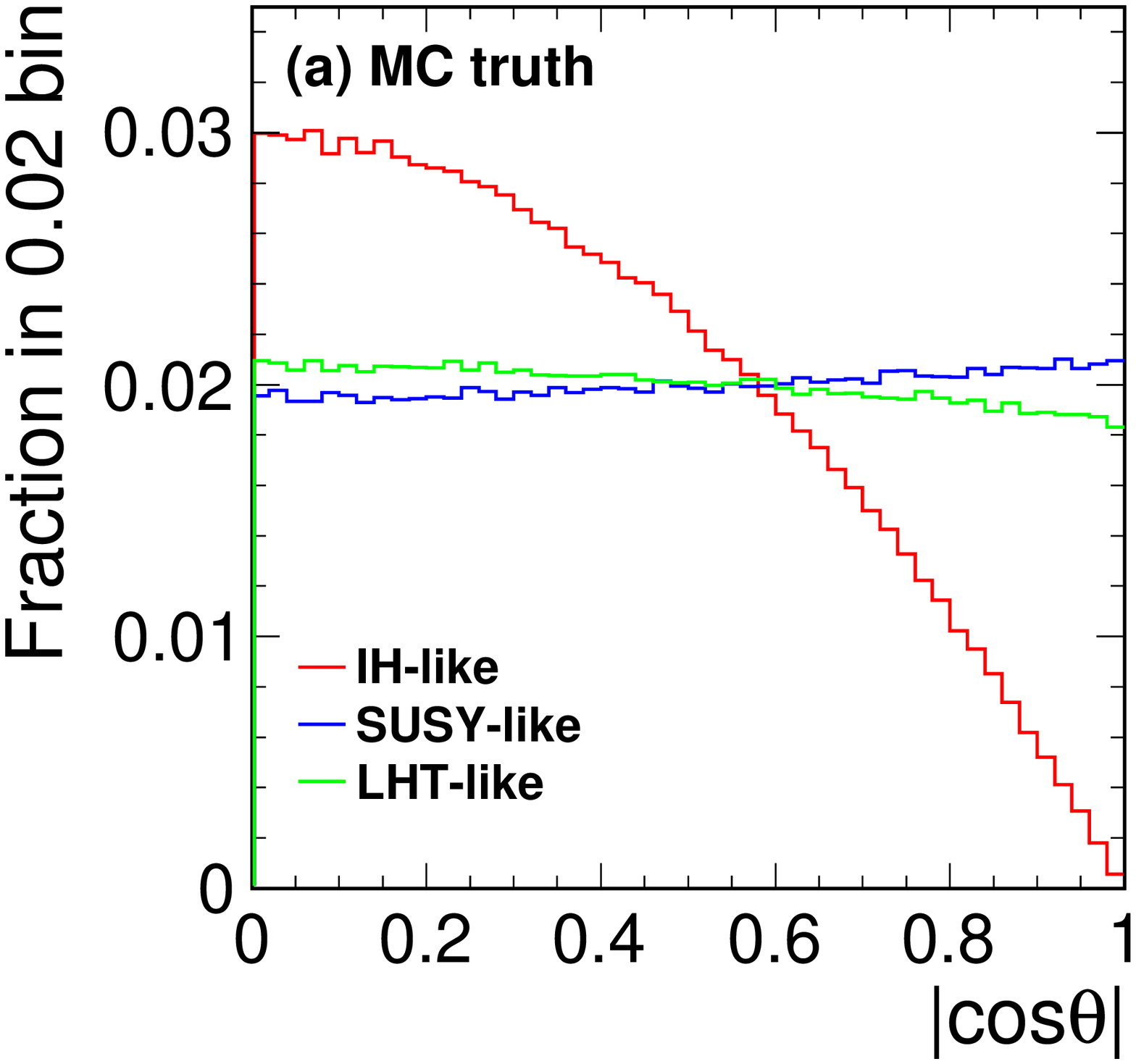}
	\end{center}
	\end{minipage}
	\begin{minipage}[t]{.48\textwidth}
	\begin{center}
    \includegraphics[width=.8\textwidth]{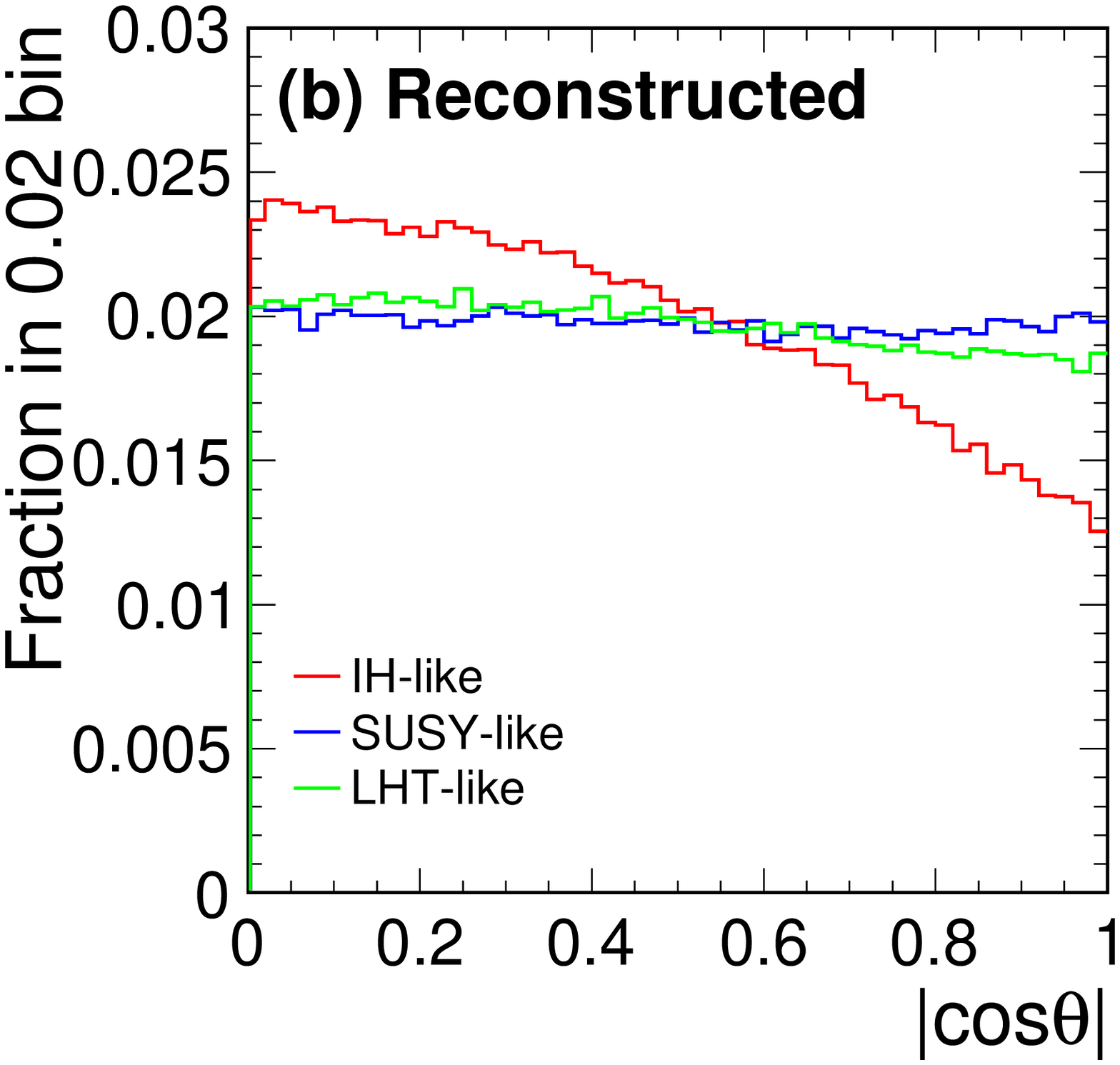}
	\end{center}
	\end{minipage}
\\
	\begin{minipage}[t]{.48\textwidth}
	\begin{center}
    \includegraphics[width=.8\textwidth]{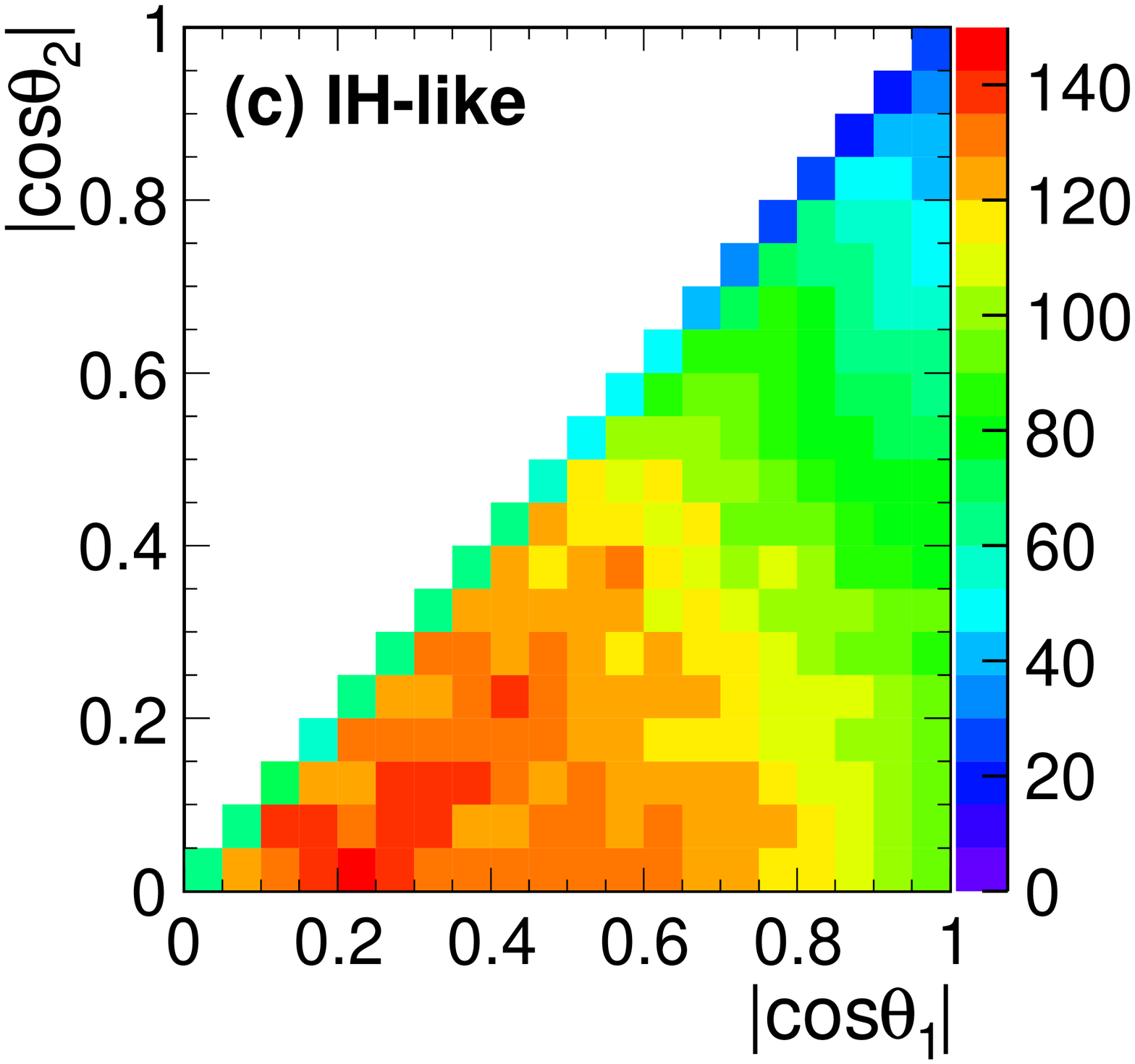}
	\end{center}
	\end{minipage}
	\begin{minipage}[t]{.48\textwidth}
	\begin{center}
    \includegraphics[width=.8\textwidth]{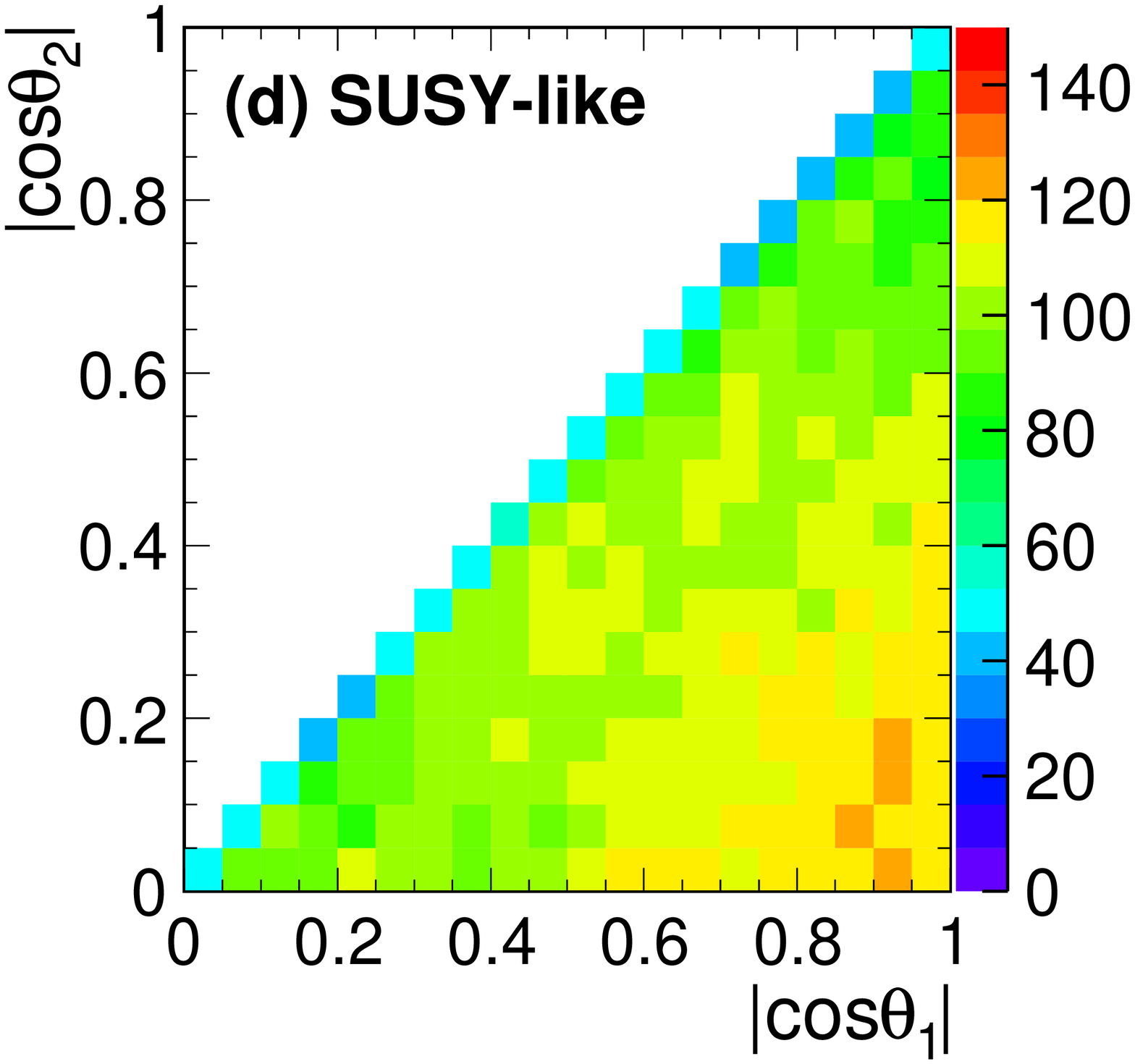}
	\end{center}
	\end{minipage}
\\
	\begin{minipage}[t]{.48\textwidth}
	\begin{center}
    \includegraphics[width=.8\textwidth]{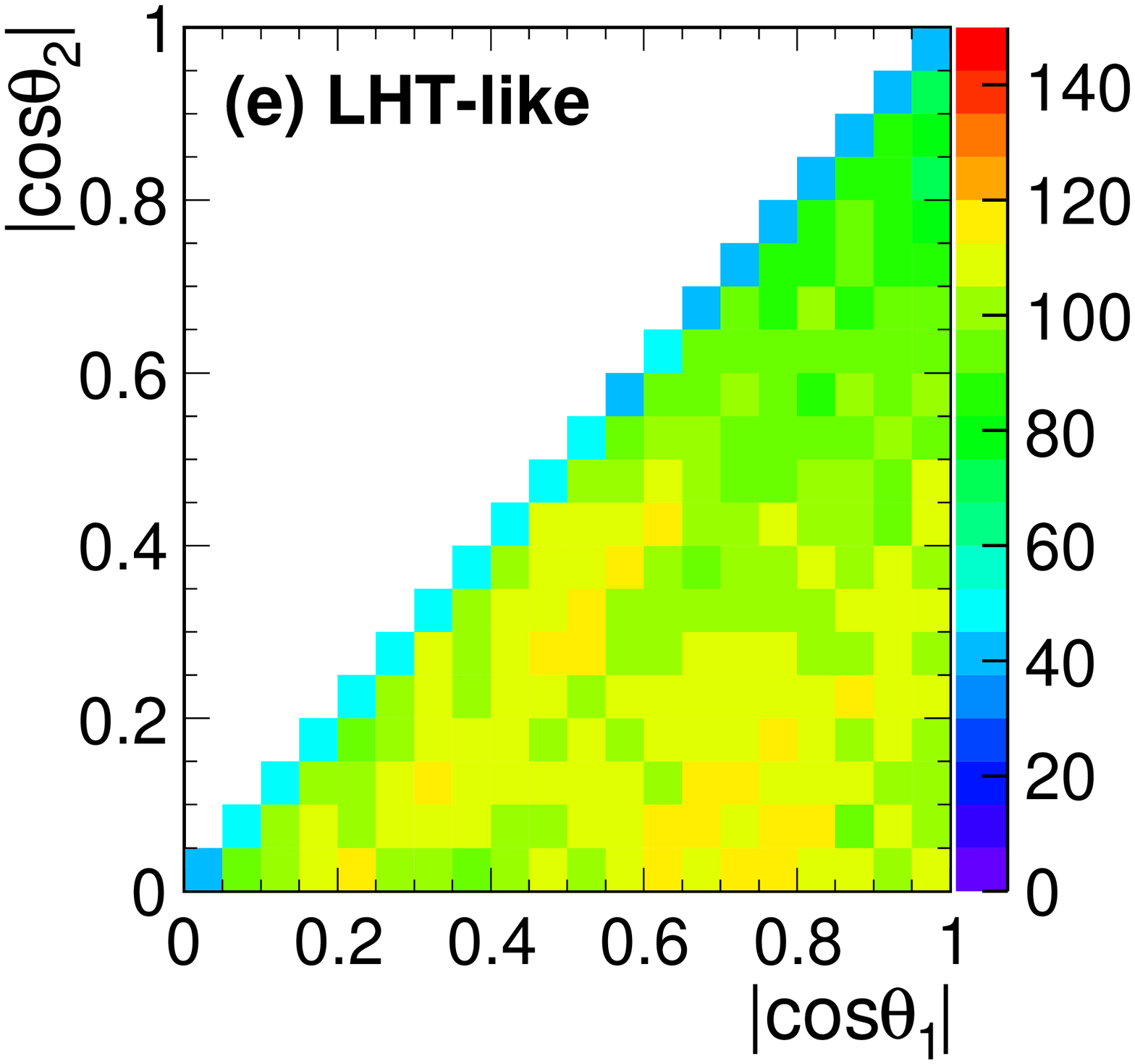}
	\end{center}
	\end{minipage}
	\begin{minipage}[t]{.48\textwidth}
	\begin{center}
    \includegraphics[width=.8\textwidth]{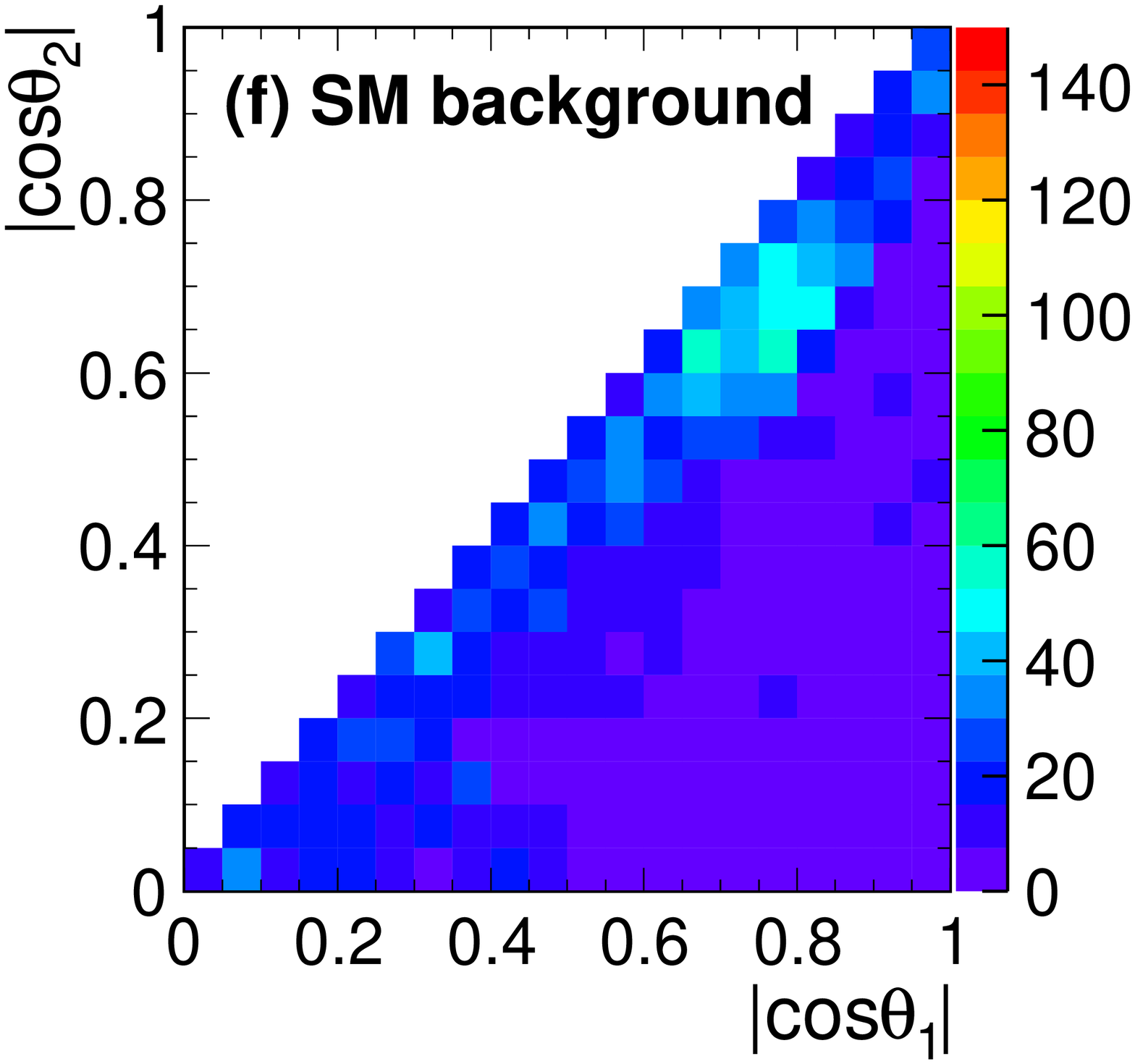}
	\end{center}
	\end{minipage}
	\end{center}
	\caption{Production angle distributions in the Point I study.
(a) and (b) show the generated and the reconstructed 1-dimensional distributions.
Both of two solutions of the quadratic equation are included in (b).
The difference between (a) and (b) reflects the effect of the wrong solution as well as the detector response.
(c)-(f) give 2-dimensional distributions of the IH-like, the SUSY-like, the LHT-like models, and the SM background, normalized to $\sigma_s = 200$ fb and $\mathcal{L}_\mathrm{int} = 500$ fb$^{-1}$.}	
	\label{fig:angle500}
\end{figure}

The separation of the three models is possible by comparing the distributions of $\chi^\pm$ production angles.
To derive the production angles,
a quadratic equation is solved using the masses of new particles
and the momenta of the $W$ bosons with the assumption of a back-to-back ejection of the $\chi^\pm$ pair.
The equation gives either two solutions which contain one correct production angle
or no solutions when the discriminant of the equation is negative.
The unphysical negative discriminant comes from misreconstructing
$W$ momenta or imperfect back-to-back condition of the two $\chi^\pm$ mainly due to initial state radiation.
Fractions of 23.9\% (IH-like), 20.8\% (SUSY-like), 23.7\% (LHT-like), and 64.4\% (SM background) of
the events have negative discriminant and are discarded before the following analysis.

\begin{figure}[p]
	\begin{center}
	\begin{minipage}[t]{.48\textwidth}
	\begin{center}
    \includegraphics[width=1\textwidth]{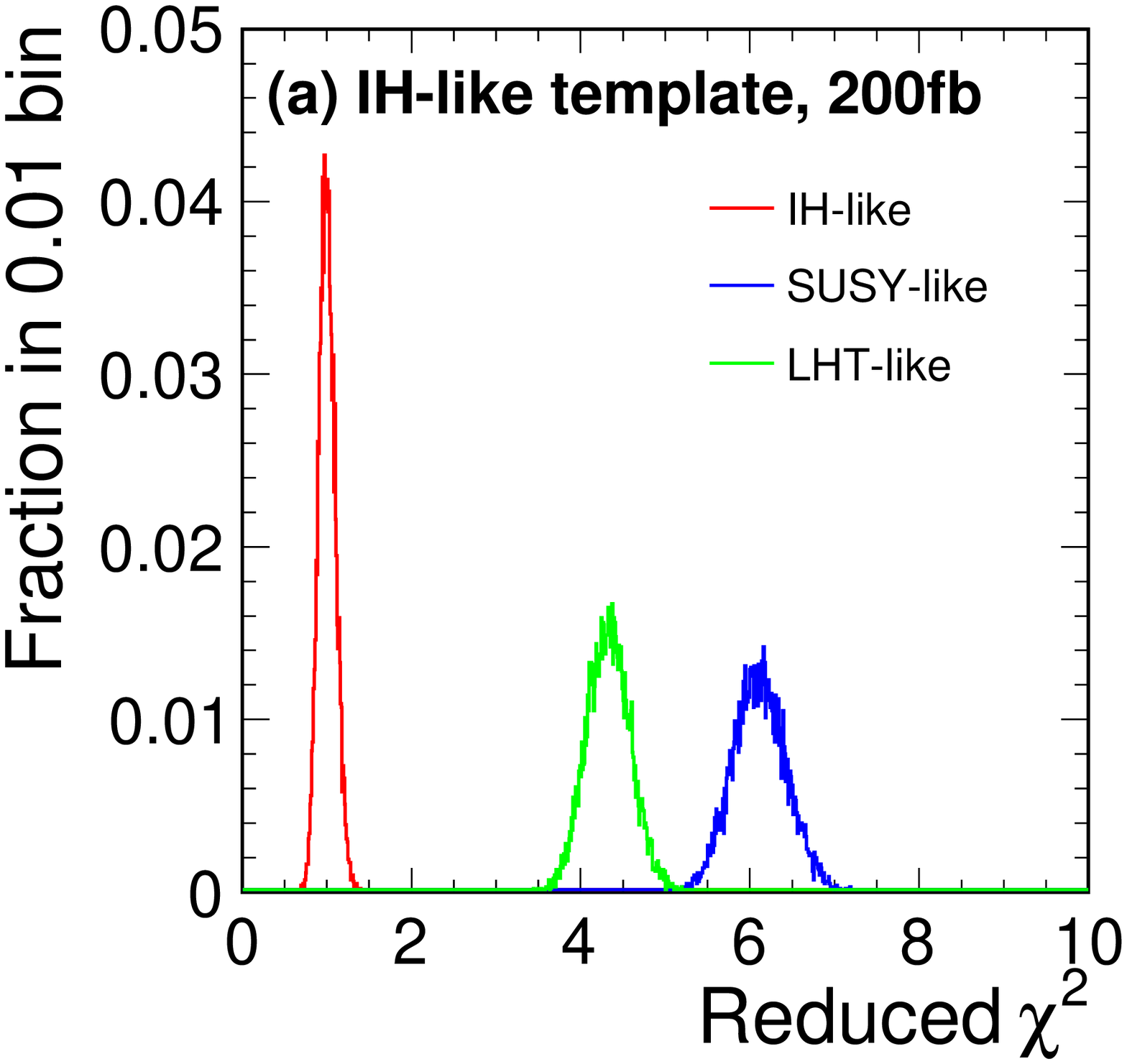}
	\end{center}
	\end{minipage}
	\begin{minipage}[t]{.48\textwidth}
	\begin{center}
    \includegraphics[width=1\textwidth]{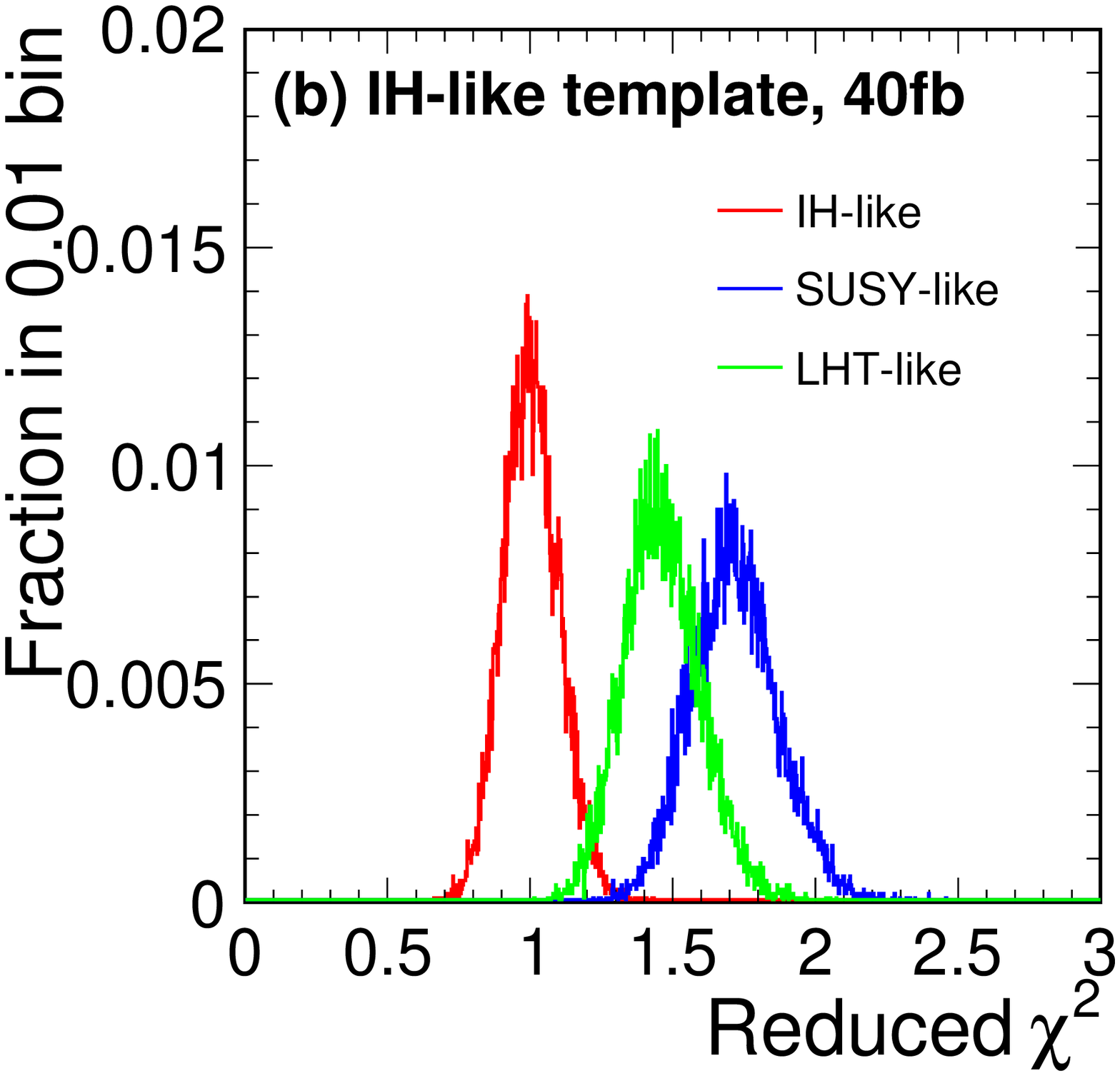}
	\end{center}
	\end{minipage}
\\
	\begin{minipage}[t]{.48\textwidth}
	\begin{center}
    \includegraphics[width=1\textwidth]{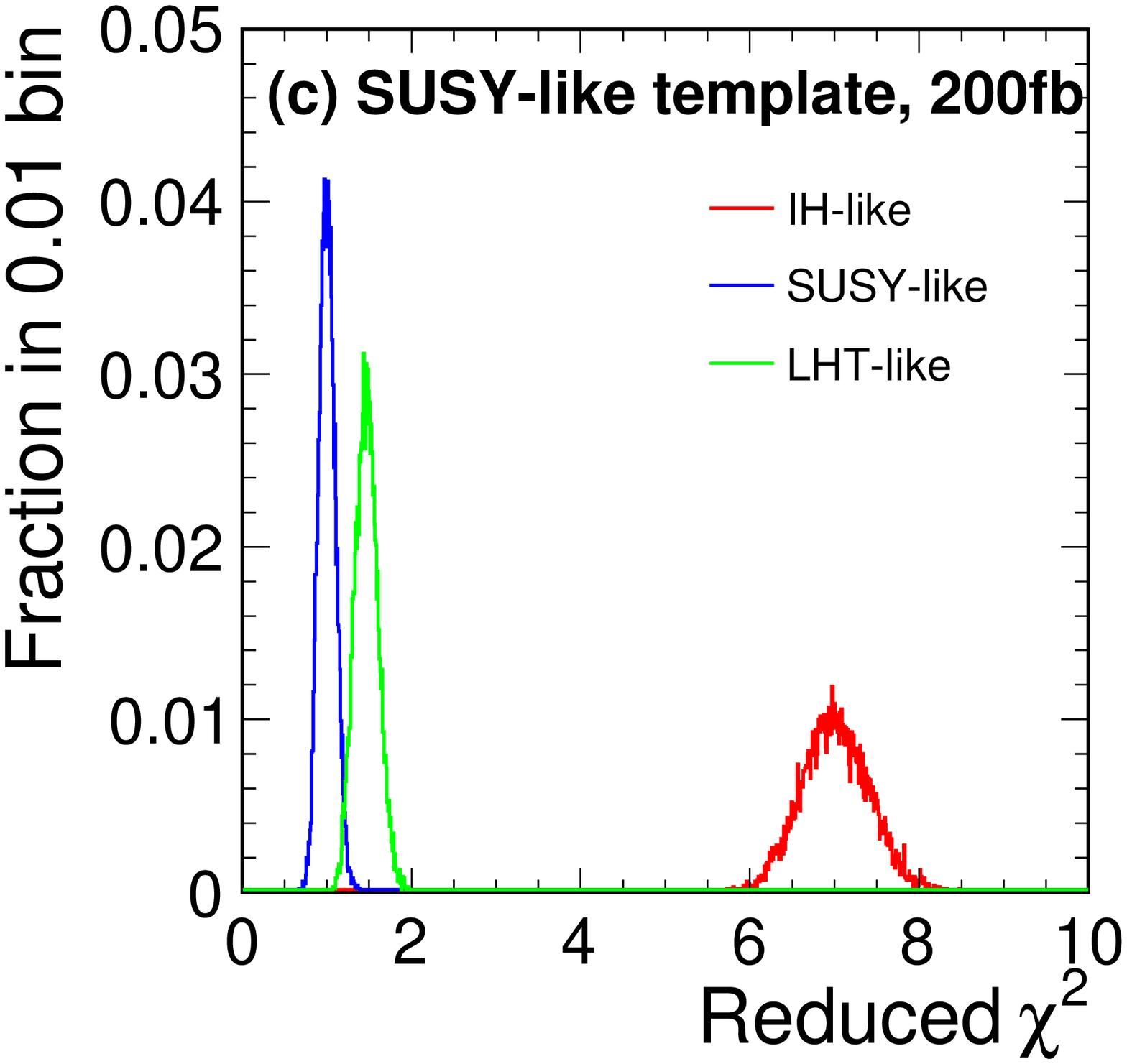}
	\end{center}
	\end{minipage}
	\begin{minipage}[t]{.48\textwidth}
	\begin{center}
    \includegraphics[width=1\textwidth]{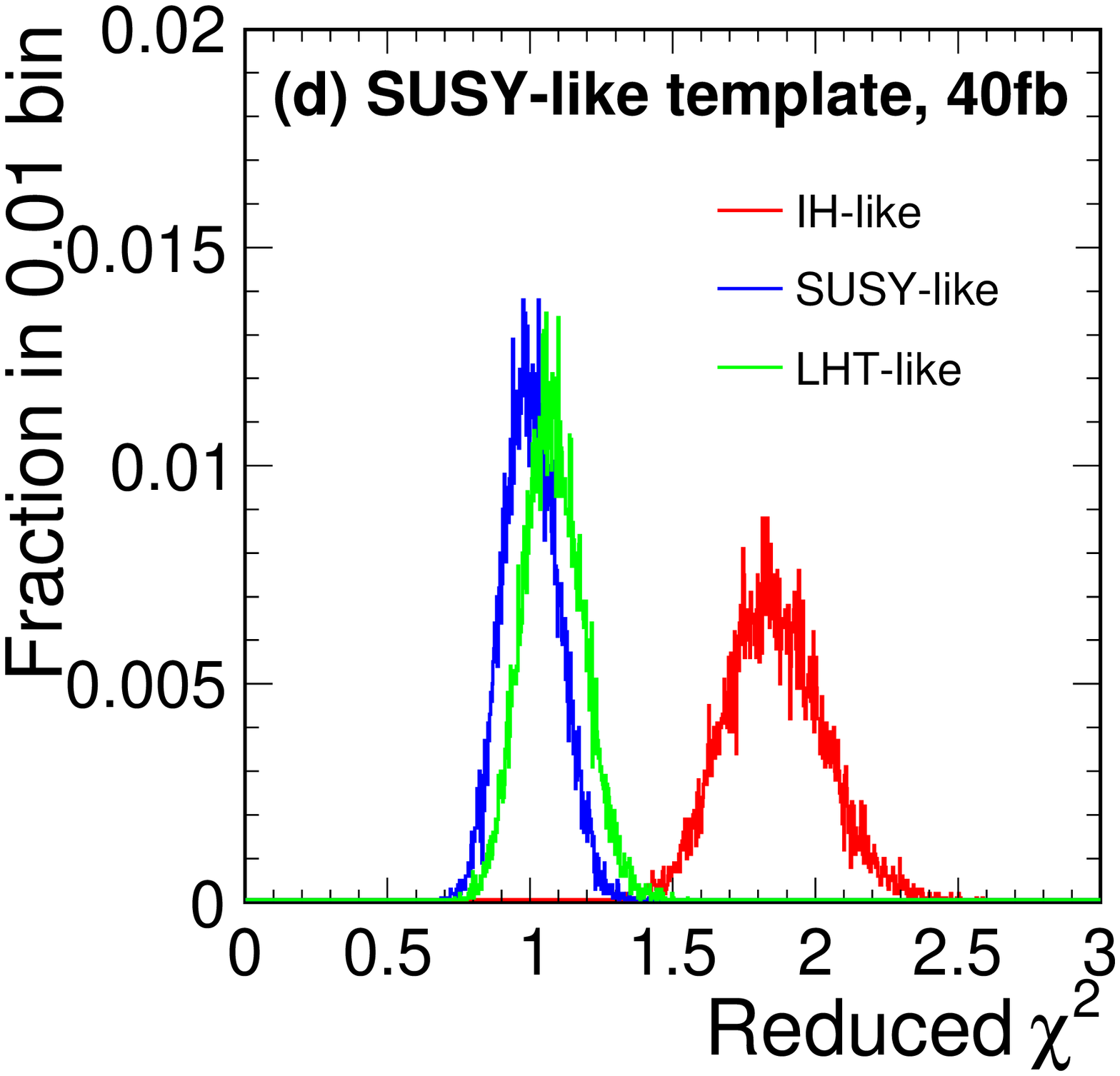}
	\end{center}
	\end{minipage}
\\
	\begin{minipage}[t]{.48\textwidth}
	\begin{center}
    \includegraphics[width=1\textwidth]{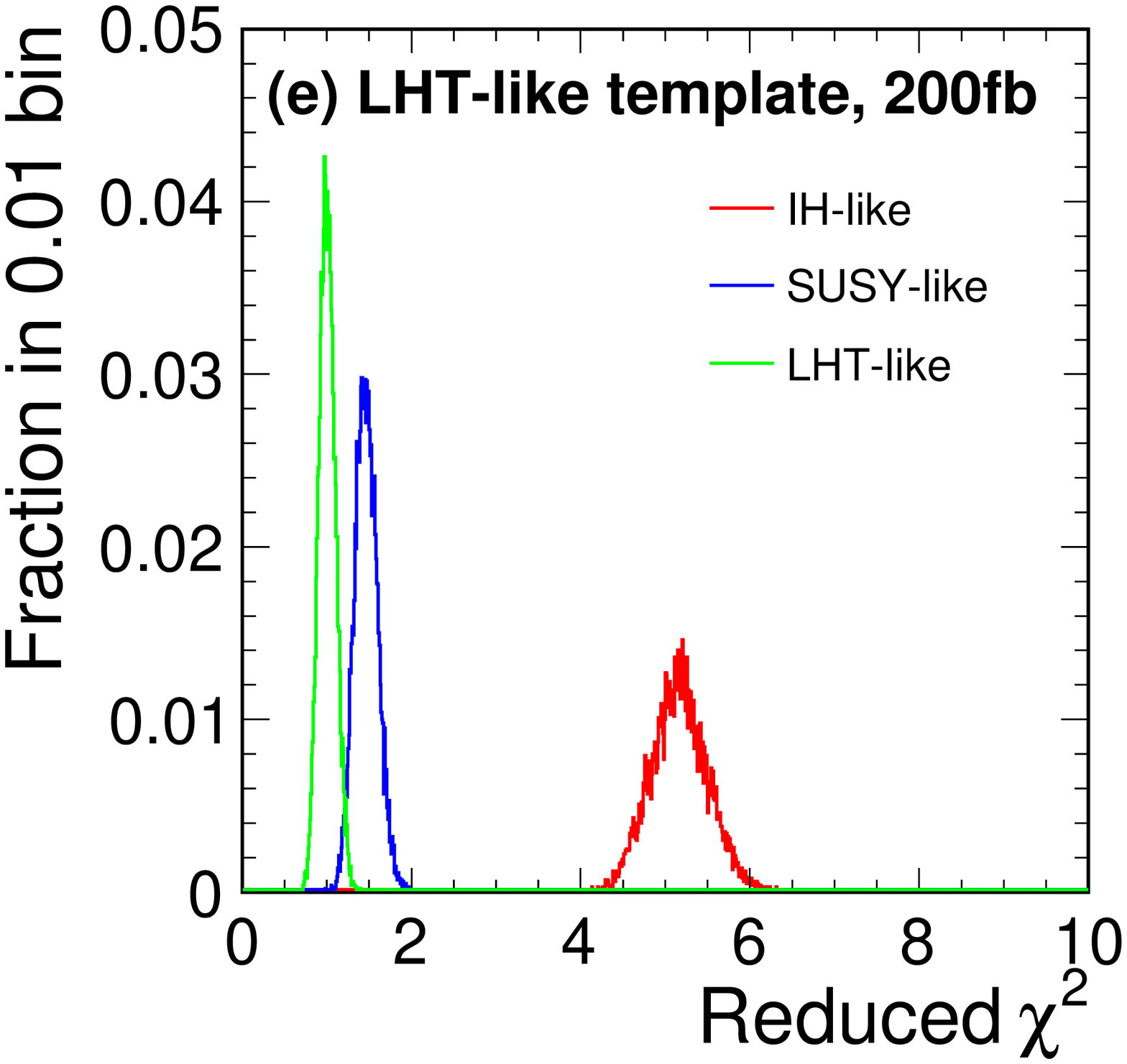}
	\end{center}
	\end{minipage}
	\begin{minipage}[t]{.48\textwidth}
	\begin{center}
    \includegraphics[width=1\textwidth]{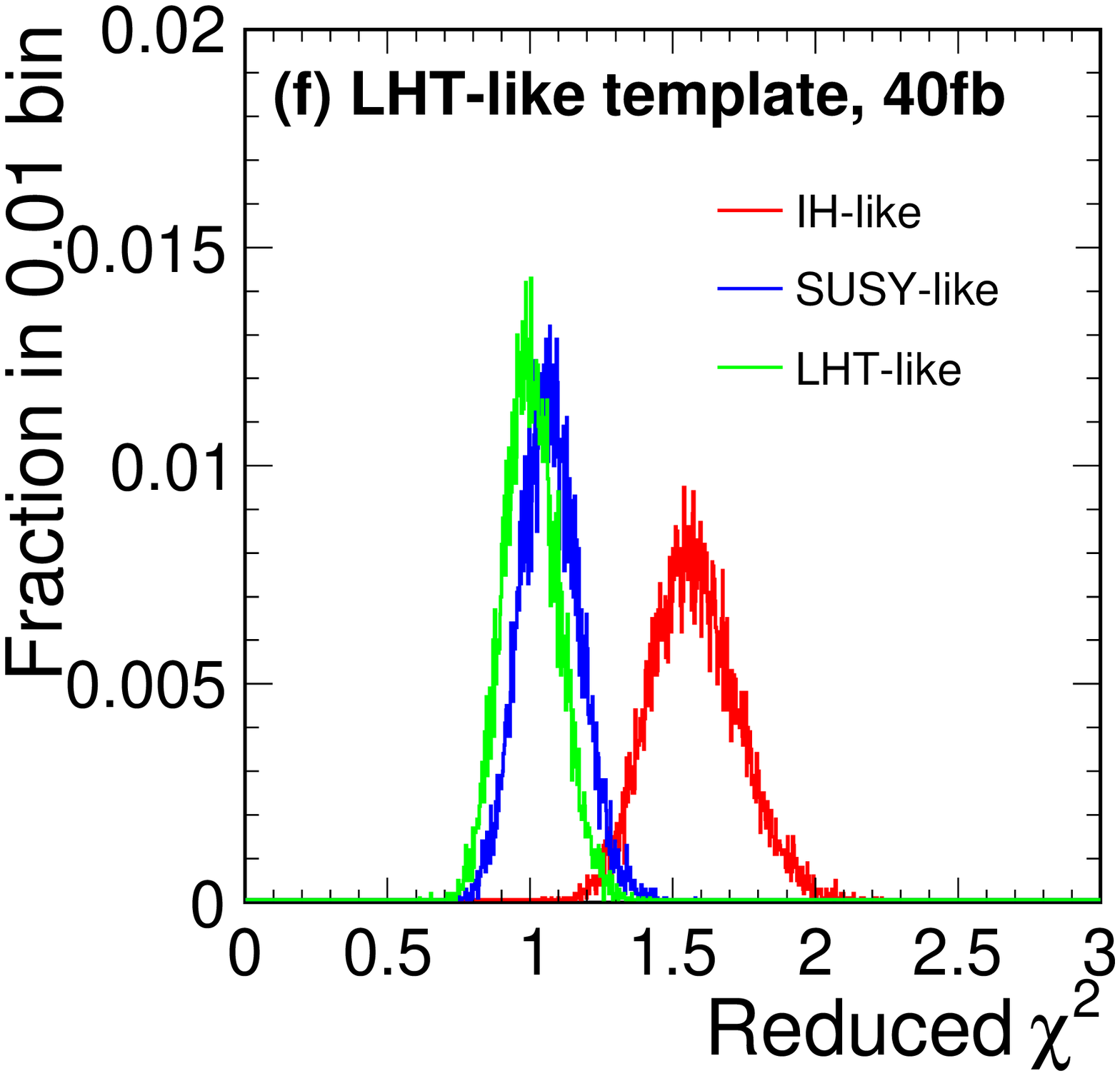}
	\end{center}
	\end{minipage}
	\end{center}
	\caption{$\tilde{\chi^2}(M_D, M_T)$ distributions with $\mathcal{L}_\mathrm{int} = 500$ fb$^{-1}$ in the Point I study.}	
	\label{fig:chi2dist}
\end{figure}

Figure \ref{fig:angle500} shows the production angle distributions.
One-dimensional results (a)(b) show the visible difference among the three models that
the IH-like events concentrate in the central region while the SUSY-like and the LHT-like events are almost flatly distributed.
Two-dimensional results (c)-(f) are actually used to estimate the separation power.
We compare the two-dimensional production angle distribution for
one model (dubbed as ``dataset'') against another model (``template'').
Distributions (c)-(e) are used as templates for each model after adding the SM background (f).
Datasets for each model are created by fluctuating each bin of the templates with Poisson distribution.
To quantify the difference between a dataset of the model $M_D$ and a template of the model $M_T$,
we defined the chi-square $\chi^2(M_D, M_T)$, the reduced chi-square $\tilde{\chi^2}(M_D, M_T)$ and
the separation power $P(M_D, M_T)$ as
\begin{eqnarray}
	\chi^2(M_D, M_T) &=& \sum_i^\mathrm{bins}\frac{\left\{D_i(M_D) - T_i(M_T)\right\}^2}{|T_i(M_T)|} \nonumber \\
	\label{eq:chi2}
	\tilde{\chi^2}(M_D, M_T) &=& \frac{\chi^2(M_D, M_T)}{N-1} \\
	P(M_D, M_T) &=& \frac{\tilde{\chi^2}(M_D, M_T) - 1}{\sigma(M_T)} \nonumber
\end{eqnarray}
where $D_i(M)$ and $T_i(M)$ are the numbers of the dataset and the template events in the $i$th bin of the model $M$,
$N = 210$ is the number of bins and $\sigma(M)$ is the standard deviation of the $\tilde{\chi^2}(M, M)$.
Since we use a high-statistics sample (1 million events for each model) for the template,
the effect of the MC statistics of the templates can be ignored.
The template distributions are normalized to the integral of the data events before calculating the $\chi^2$ value.
Figure \ref{fig:chi2dist} shows the obtained $\tilde{\chi^2}$ distribution with 10,000 datasets for every combination of the three models.
Separation is possible for every model with $\sigma_s = 200$ fb, while
in the $\sigma_s = 40$ fb case clear separation between the SUSY-like and the LHT-like models is impossible. 

\begin{table}
\center{
\begin{tabular}{|l|r|r|r|r|}
\hline
$\sigma_s$  & $M_D \backslash M_T$  & IH-like & SUSY-like & LHT-like \\
\hline
200 fb &  IH-like  &  - & 63  & 43 \\
       & SUSY-like & 53 & -   & 4.9 \\
       & LHT-like  & 35 & 4.9 & - \\ \hline
40 fb  & IH-like   & -   & 8.9 & 6.0 \\
       & SUSY-like & 7.5 & -   & 0.7 \\
       & LHT-like  & 4.9 & 0.8 & - \\ 
\hline
\end{tabular}
}
\caption{Expectation value of separation power $\bar{P}$ between the three models with the 2-dimensional production angle distribution with $\mathcal{L}_\mathrm{int} = 500$ fb$^{-1}$ in the Point I study.}
 \label{tbl:prodangle}
\end{table}

Table \ref{tbl:prodangle} tabulates the expected values of obtained separation power $\bar{P}$.
Despite the similar angular distribution of the SUSY-like and the LHT-like models,
all the three models can be identified with $\sigma_s = 200$ fb.
In the $\sigma_s = 40$ fb case, the SUSY-like and the LHT-like models cannot be separated
while the IH-like model can still be separated from the other two.
These values do not include the effect of the mass uncertainty of new particles,
which is not significant with $<$ 5\% mass uncertainty obtained in our mass determination analysis (see Table \ref{tbl:massfit500}).


\subsubsection{Threshold Scan}

\begin{figure}
	\begin{center}
		\begin{minipage}{.32\textwidth}
			\begin{center}
				\includegraphics[width=1\textwidth]{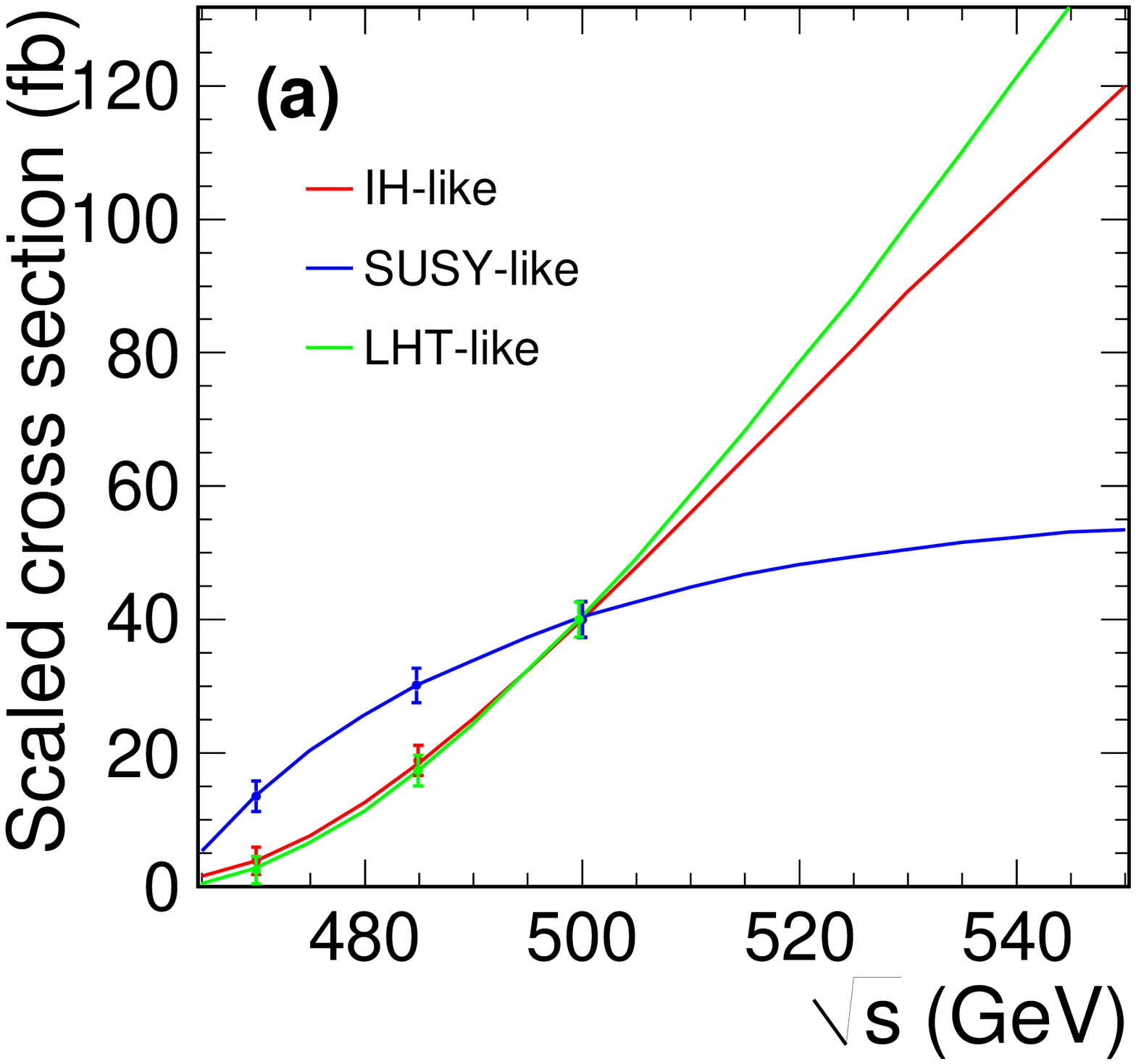}
			\end{center}
		\end{minipage}
		\begin{minipage}{.32\textwidth}
			\begin{center}
				\includegraphics[width=1\textwidth]{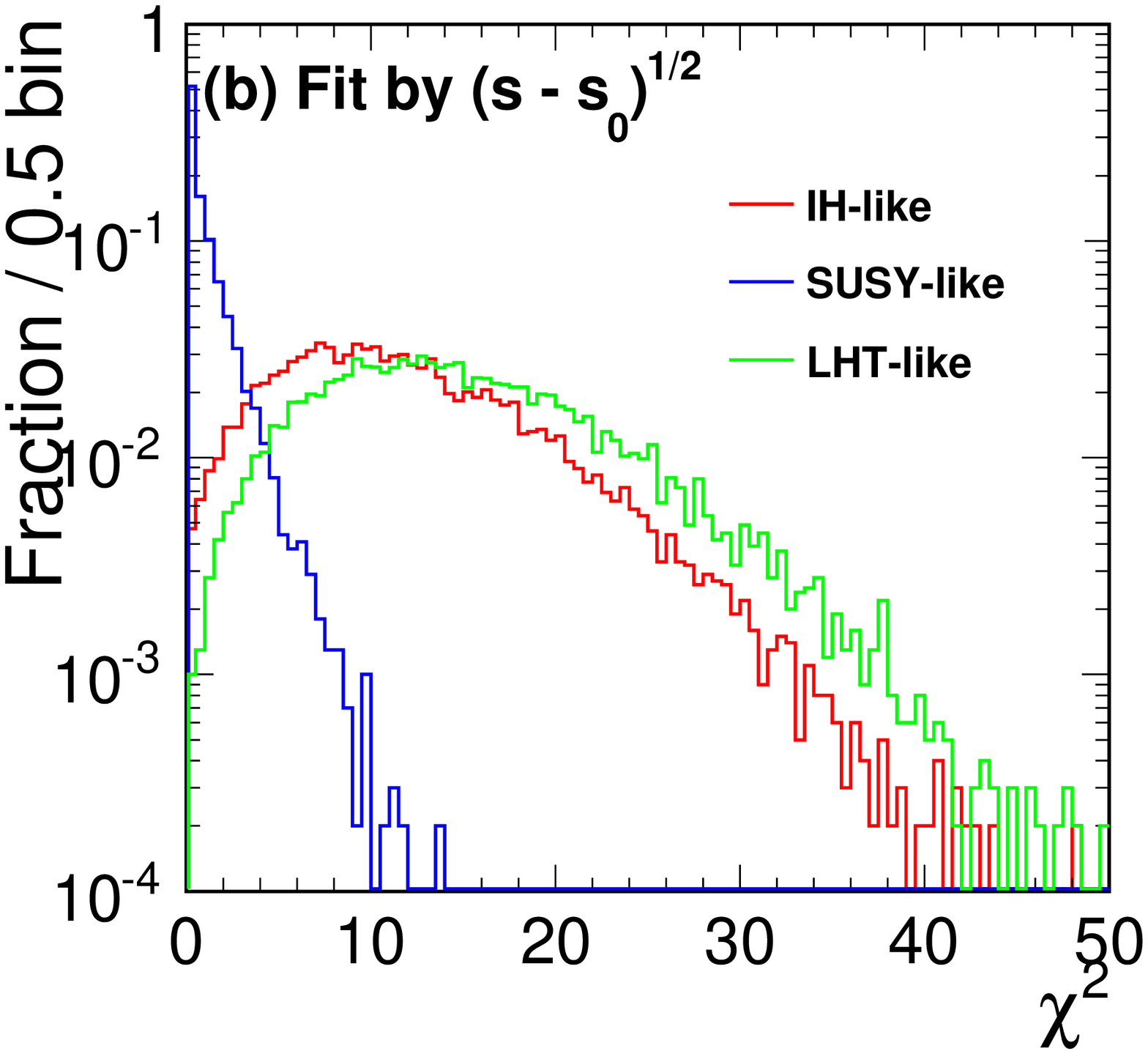}
			\end{center}
		\end{minipage}
		\begin{minipage}{.32\textwidth}
			\begin{center}
				\includegraphics[width=1\textwidth]{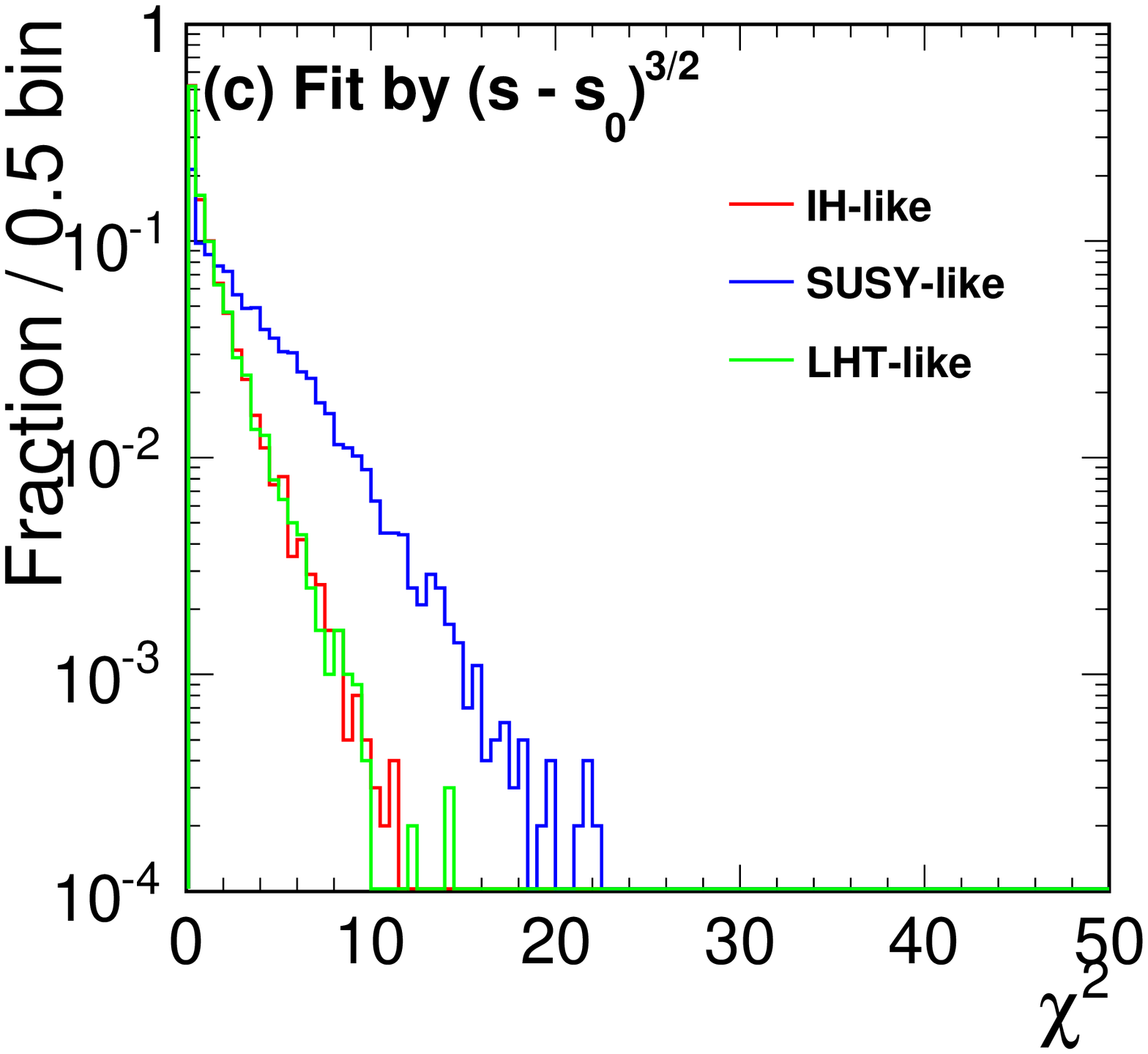}
			\end{center}
		\end{minipage}
	\end{center}
	\caption{(a) Dependence of the cross section on the center of mass energy, normalized to $\sigma_s = 40$ fb in the Point I study. Error bars are given assuming $\mathcal{L}_\mathrm{int} = 50$ fb$^{-1}$ data at each point.  (b),(c) Results of the $\chi^2$ fits for the $(s - s_0)^{1/2}$ case and the $(s - s_0)^{3/2}$ case, respectively.}
\label{fig:thscan}
\end{figure}

Another strategy to distinguish the models is the threshold scan.
Figure \ref{fig:thscan} (a) shows how the cross section of each model depends on $\sqrt{s}$.
A clear difference can be seen between the SUSY-like model whose production cross section has the $(s - s_0)^{1/2}$ dependence
and the other two whose cross sections have the $(s - s_0)^{3/2}$ dependence
where $s_0$ is the threshold energy, which is twice the mass of the $\chi^\pm$.

To estimate the separation power with the threshold scan, we performed a toy-MC study, in which the measured cross section
was fluctuated using the expected signal and background statistics obtained from the full-MC study.
The cut efficiency and the background cross section are assumed to be identical to the 500 GeV case for any $\sqrt{s}$ .
We performed a scan of three points: $\sqrt{s} =$ 470, 485, and 500 GeV, each with $\mathcal{L}_\mathrm{int} = 50$ fb$^{-1}$.
The cross section is scaled so that all the three models have $\sigma_s = 40$ fb at 500 GeV.

For the separation, we calculate the $\chi^2$ value of the fit of
\begin{equation}
	\sigma(s, n) = a(s - s_0)^{n}, \qquad n = 1/2, \qquad 3/2
	\label{eq:threshold}
\end{equation}
where $a$ and $s_0$ are the free parameters for each model.
Figures \ref{fig:thscan} (b) and (c) show the $\chi^2$ distributions. With the $n = 1/2$ fit (b),
good separation is obtained between the SUSY-like and the other two models. For example, 92.0\% of the SUSY-like events are within 
$\chi^2 < 3$ while 5.7 and 2.1\% of the IH-like and LHT-like events remain in the same $\chi^2$ region.
The $n = 3/2$ fit (c) does not have significant separation power.
Separation between the IH-like and the LHT-like models is almost impossible by the threshold scan.

Since the SUSY-like model can be separated from the LHT-like model with the threshold scan and the IH-like model can be separated
from the SUSY-like model with the production angle distribution, the three models can be separated from each other
with combining the two methods even in the $\sigma_s = 40$ fb case.

%% file: Analysis1000.tex
\subsection{Study for Point II with $\sqrt{s} = 1$ TeV fast simulation}

Since most of the analysis procedure is the same as in the Point I study,
we mainly focus on the difference and the result of the Point II study in this subsection.

\subsubsection{Signal selection}
Point II ($m_{\chi^\pm} = 368$ GeV, $m_{\chi^{0}} =  81.9$ GeV) is not accessible with $\sqrt{s} = 500$ GeV ILC,
so we use 1 TeV fast simulation for the Point II study.
 As in the Point I study, hadronic decay modes of $W$ bosons have been used to select the signal process. All events were reconstructed as 4-jet events by adjusting the cut on $y$-values. In order to identify the two $W$ bosons from $\chi^{\pm}$ decays, two jet-pairs were selected so as to minimize a $\chi^2$ function,
\begin{equation}
\chi^2
= 
(^{\mathrm{rec}}M_{W1} -~^{\mathrm{tr}}M_{W})^{2}/\sigma_{M_{W}}^{2} 
+ 
(^{\mathrm{rec}}M_{W2} -~^{\mathrm{tr}}M_{W})^{2}/\sigma_{M_{W}}^{2},
\end{equation}
where $^{\mathrm{rec}}M_{W1(2)}$ is the invariant mass of the first (second) 2-jet system paired as a $W$ candidate, $^{\mathrm{tr}}M_{W}$ is the true $W$ mass (80.4 GeV), and $\sigma_{M_{W}}$ is the resolution for the $W$ mass (4 GeV). We required $\chi^2 < 26$ to obtain well-reconstructed events. 
Since $\chi^{0}$'s escape from detection resulting in missing momentum, the missing transverse momentum (\misspt) of the signal peaks at around 175 GeV. We have thus selected events with \misspt~above 84 GeV. The numbers of events after the selection cuts are summarized in Table \ref{tb:cut_summary1tev}.
Leptonic decay in $\chi^\pm$ pair production and SM Higgs backgrounds are not included in the Point II study.
These backgrounds are expected to be small according to the Point I study.

\subsubsection{Mass determination}

\begin{figure}[p]
\begin{center}
\includegraphics[width=1\textwidth]{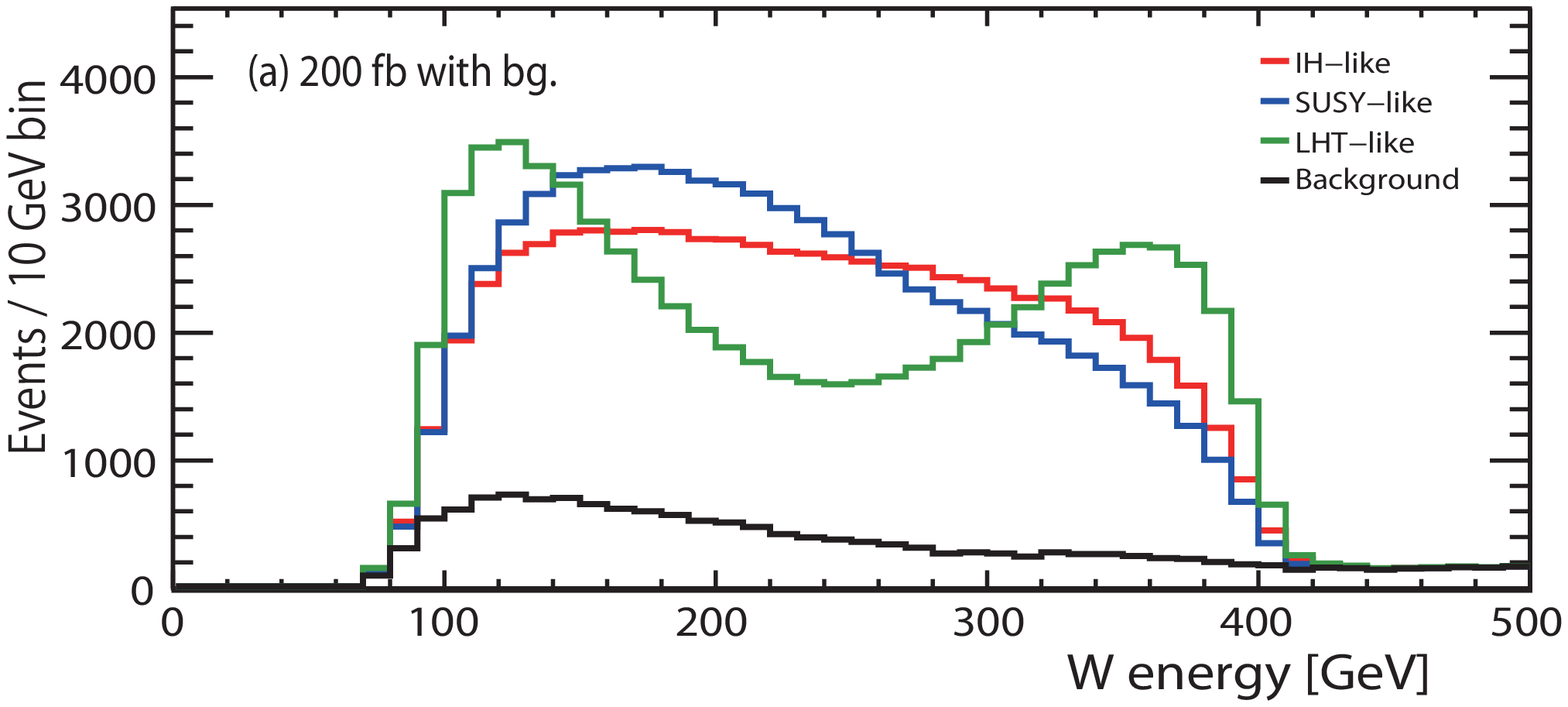}
\includegraphics[width=1\textwidth]{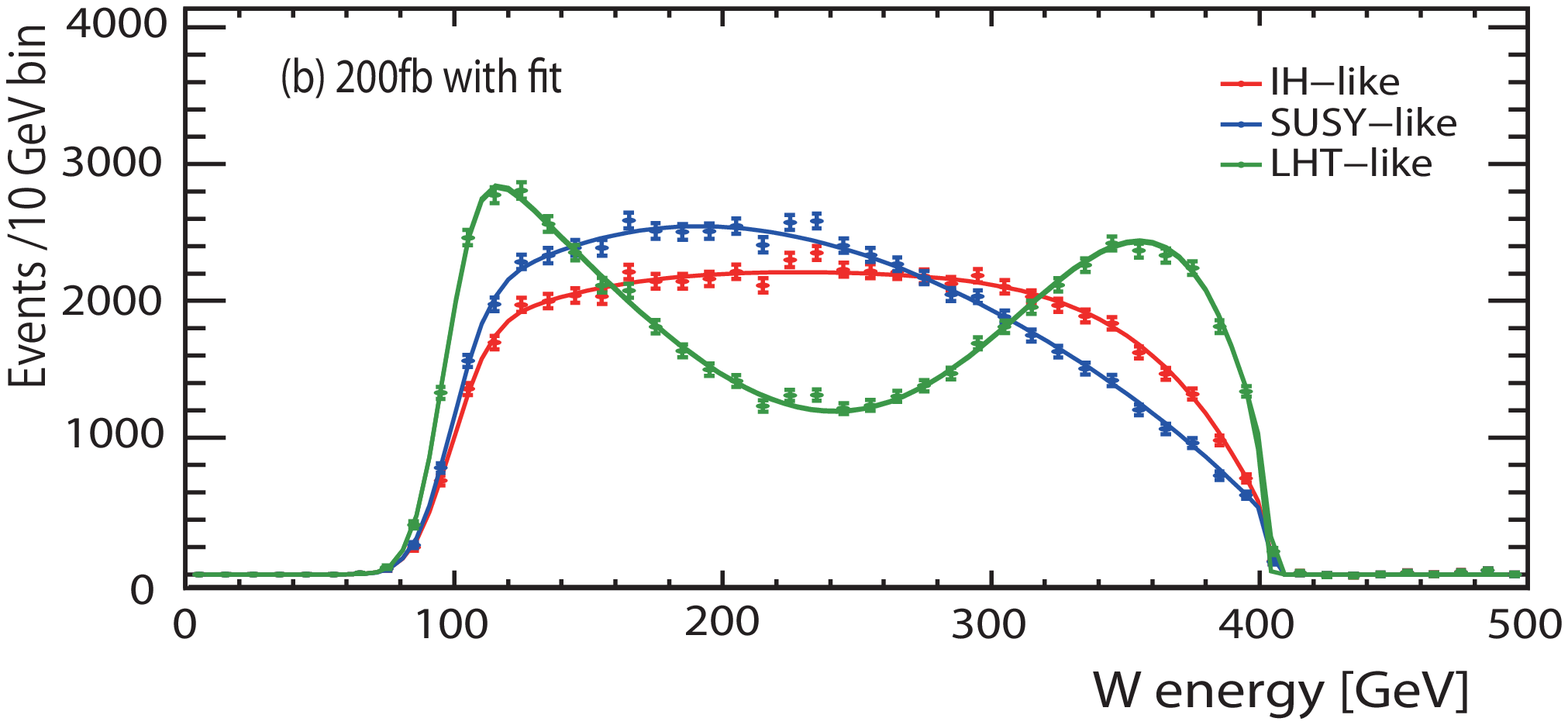}
\includegraphics[width=1\textwidth]{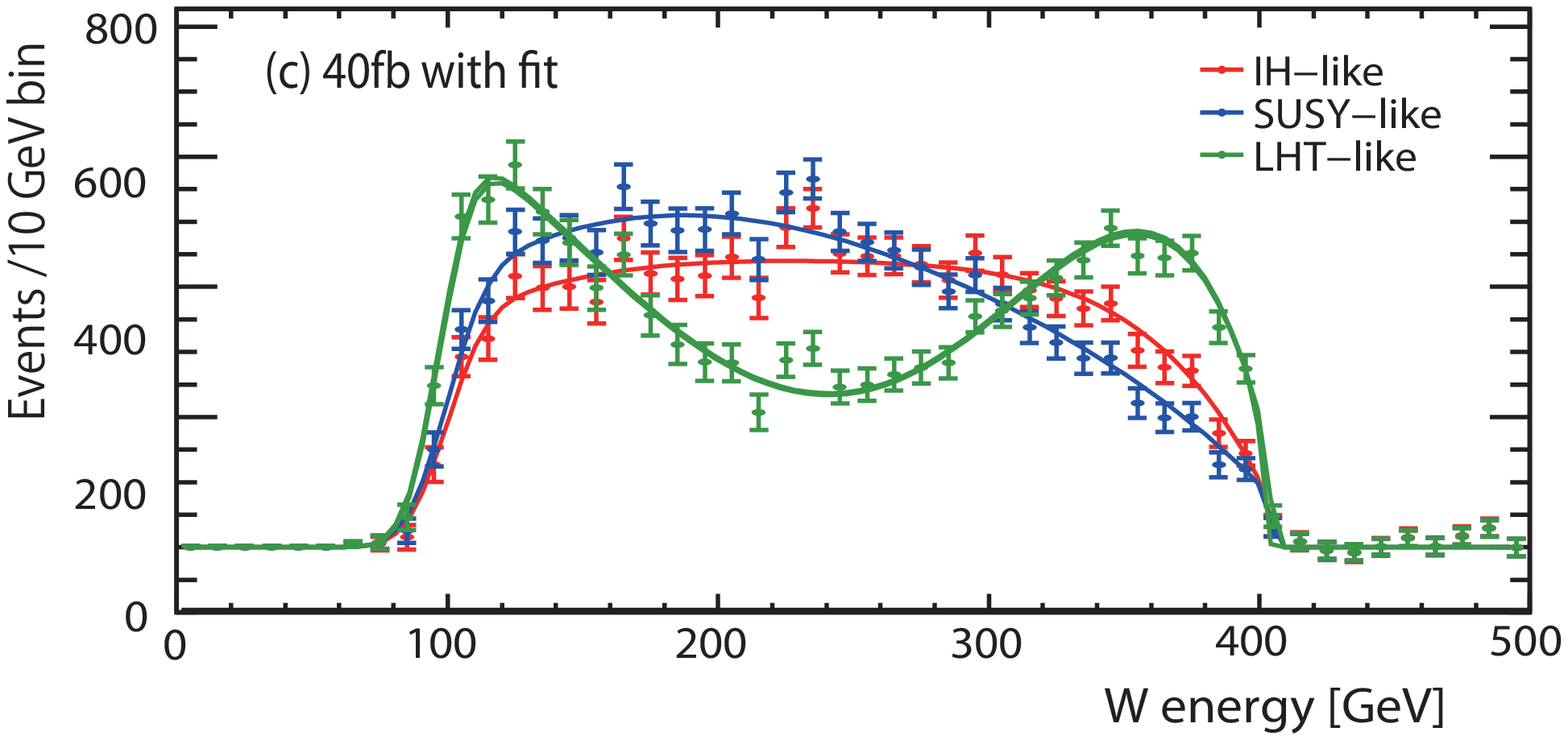}
\end{center}
\caption{(a) $W$ energy distributions for the signal ($\sigma_s = 200$ fb) and background with $\mathcal{L}_\mathrm{int} = 500$ fb$^{-1}$
		in the Point II study.
		(b),(c) Results of the mass fit for $\sigma_s = 200$ fb and 40 fb after background subtraction, respectively.
}
\label{fig:wene1tev}
\end{figure}

Procedure of the mass determination is almost the same as in the Point I study.
The masses of $\chi^{0}$ and $\chi^{\pm}$ were determined from the edges of the $W$ energy distribution shown in Fig.~\ref{fig:wene1tev}. After subtracting the backgrounds, the distribution was fitted with a line shape determined by a high statistics signal sample. The fitted masses of $\chi^{0}$ and $\chi^{\pm}$
with $\mathcal{L}_\mathrm{int} = 500$ fb$^{-1}$ are summarized in Table \ref{tb:reso_1tev}.
The masses of $\chi^\pm$ and $\chi^0$ are obtained with accuracies of better than 0.3\% and 1.5\%, respectively, for $\sigma_s = 200$ fb. 
For $\sigma_s = 40$ fb, the measurement accuracies of $\chi^\pm$ and $\chi^0$ are 0.5-1\% and 3-6\%, respectively.

\begin{table}
\center{
\begin{tabular}{|c|l|r|r|}
\hline
				& Process							& \# of events & \# of events after cuts \\
\hline
				& \ccqqqqnn (IH-like)				&  45,970 & 29,655 \\
Signal	&	\ccqqqqnn (SUSY-like)					&  45.970 & 30,335 \\
				&	\ccqqqqnn (LHT-like)			&  45,970 & 29,496 \\ \hline 
				& \enwzenqqqq						&  10,321 &  3,306 \\
				& $WWZ \to all$                     & 31,300  & 2,176 \\
				& \nnwwnnqqqq						&   3,225 &  1,473 \\			
SM bkg. &$\nu \nu ZZ \to \nu \nu qqqq$	            &   1,399 &    578 \\
				& \wwqqqq							& 886,500 &   307 \\
				& $ZZ \to qqqq$						&  67,100 &    259 \\
				& \eewweeqqqq						& 232,500 &     25 \\
\hline
\end{tabular}
}
\caption{The number of events before and after the selection cuts, normalized to $\mathcal{L}_\mathrm{int} = 500$ fb$^{-1}$
and $\sigma_s = 200$ fb in the Point II study.}
 \label{tb:cut_summary1tev}
\end{table}

\begin{table}
\center{
\begin{tabular}{|l|r|r|r|}
\hline
 & Physics model & $\sigma_s = 200$ fb & $\sigma_s = 40$ fb \\
\hline
$M_{\chi^{\pm}}$ (GeV) & IH-like   & 367.4 $\pm$ 0.9 & 366.5 $\pm$ 3.4 \\
                       & SUSY-like & 368.5 $\pm$ 0.8 & 370.7 $\pm$ 2.8 \\
                       & LHT-like  & 367.5 $\pm$ 0.6 & 367.2 $\pm$ 2.0 \\ \hline 
$M_{\chi^{0}}$ (GeV)   & IH-like   & 81.2 $\pm$ 1.1 & 80.5 $\pm$ 4.7 \\
                       & SUSY-like & 81.6 $\pm$ 1.1 & 82.5 $\pm$ 4.5 \\
                       & LHT-like  & 82.1 $\pm$ 0.8 & 84.0 $\pm$ 2.7 \\ 
\hline
\end{tabular}
}
\caption{Measurement accuracies for the masses of $\chi^{\pm}$ and $\chi^{0}$ with $\mathcal{L}_\mathrm{int} = 500$ fb$^{-1}$ in the Point II study.}
 \label{tb:reso_1tev}
\end{table}

\subsubsection{Angular distribution for $\chi^{\pm}$ pair production}

\begin{figure}[p]
\begin{center}
\begin{minipage}{.45\textwidth}
\begin{center}
\includegraphics[width=1\textwidth]{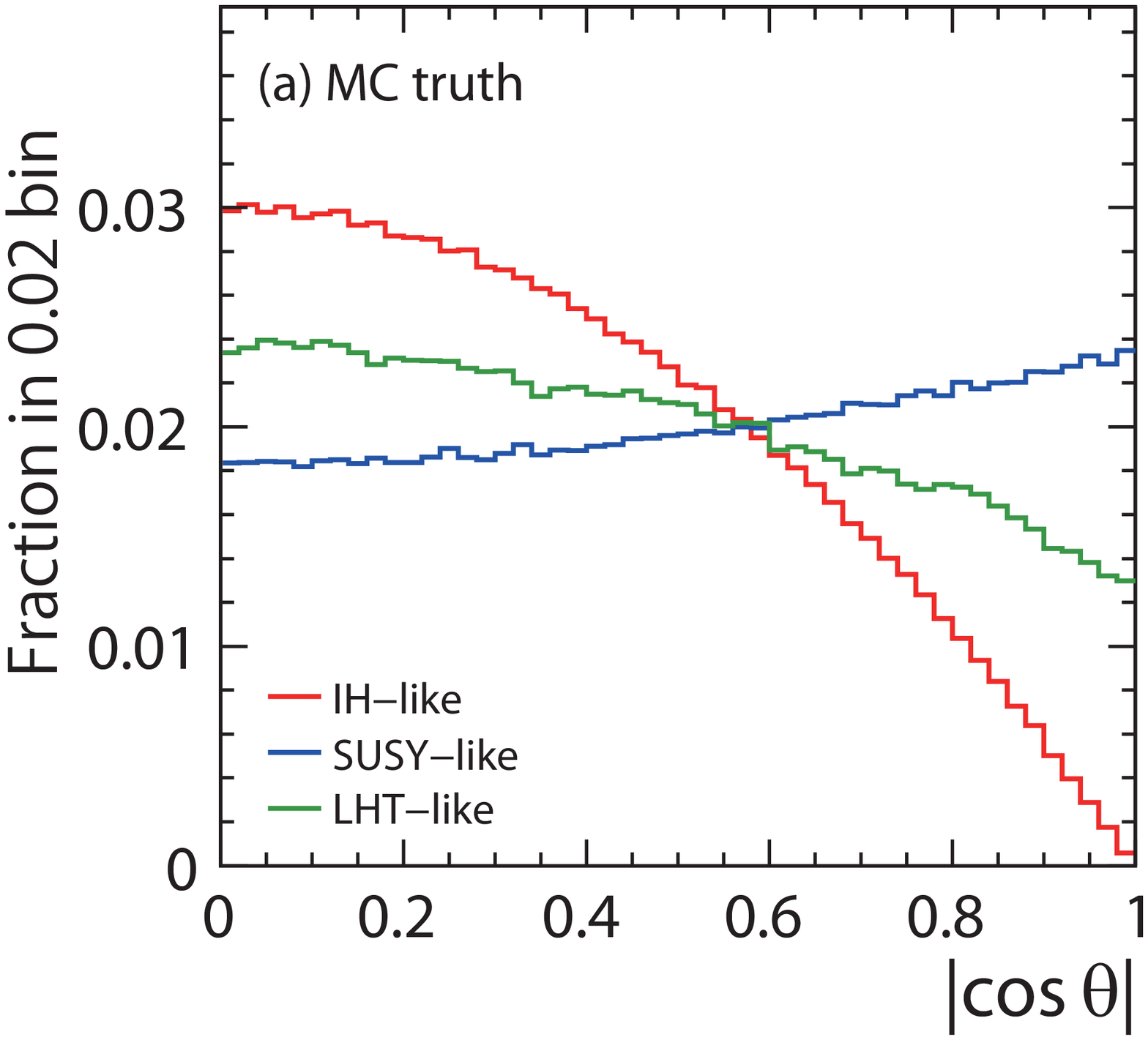}
\end{center}
\end{minipage}
\begin{minipage}{.45\textwidth}
\begin{center}
\includegraphics[width=1\textwidth]{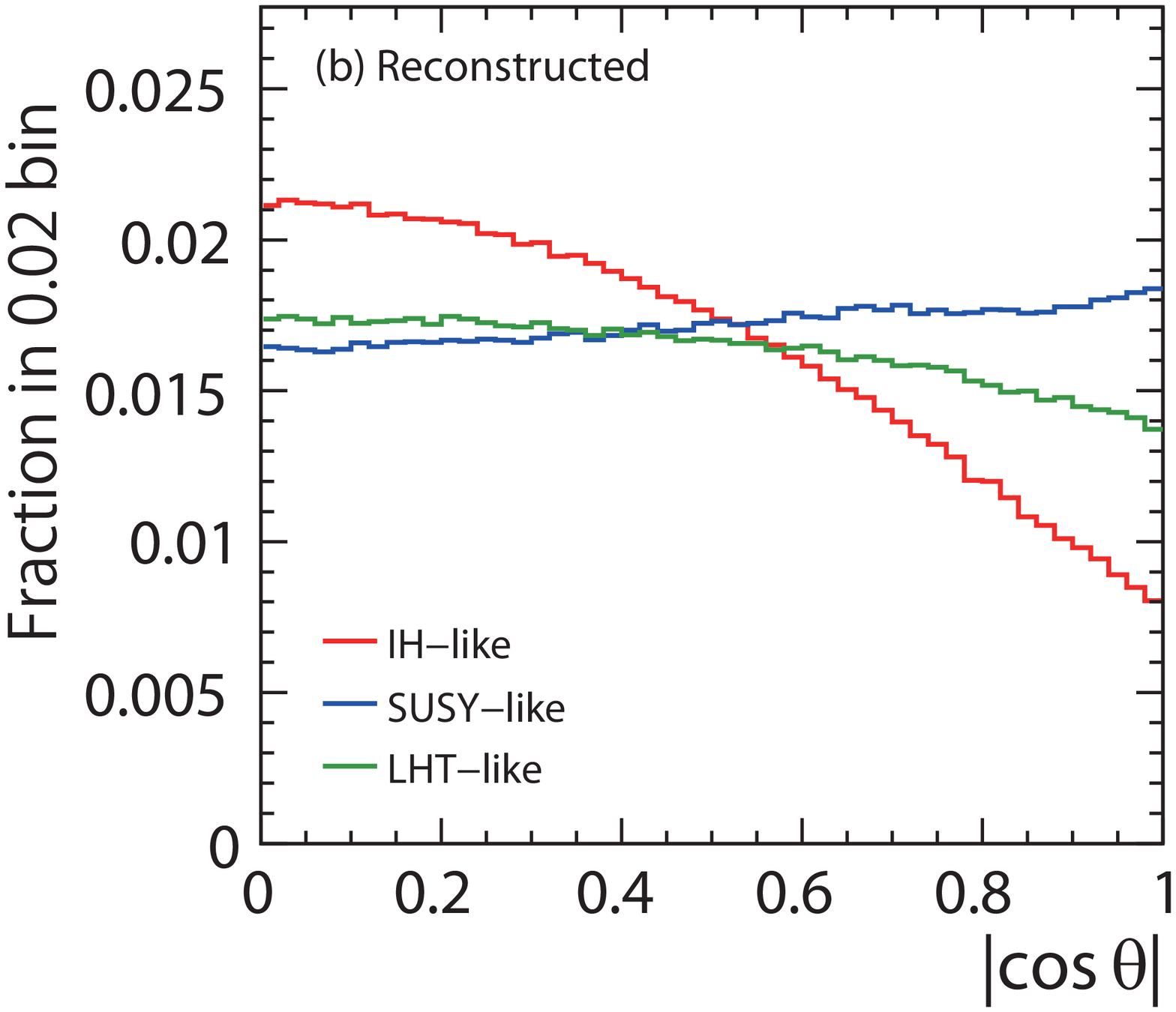}
\end{center}
\end{minipage}
\begin{minipage}{.45\textwidth}
\begin{center}
\includegraphics[width=1\textwidth]{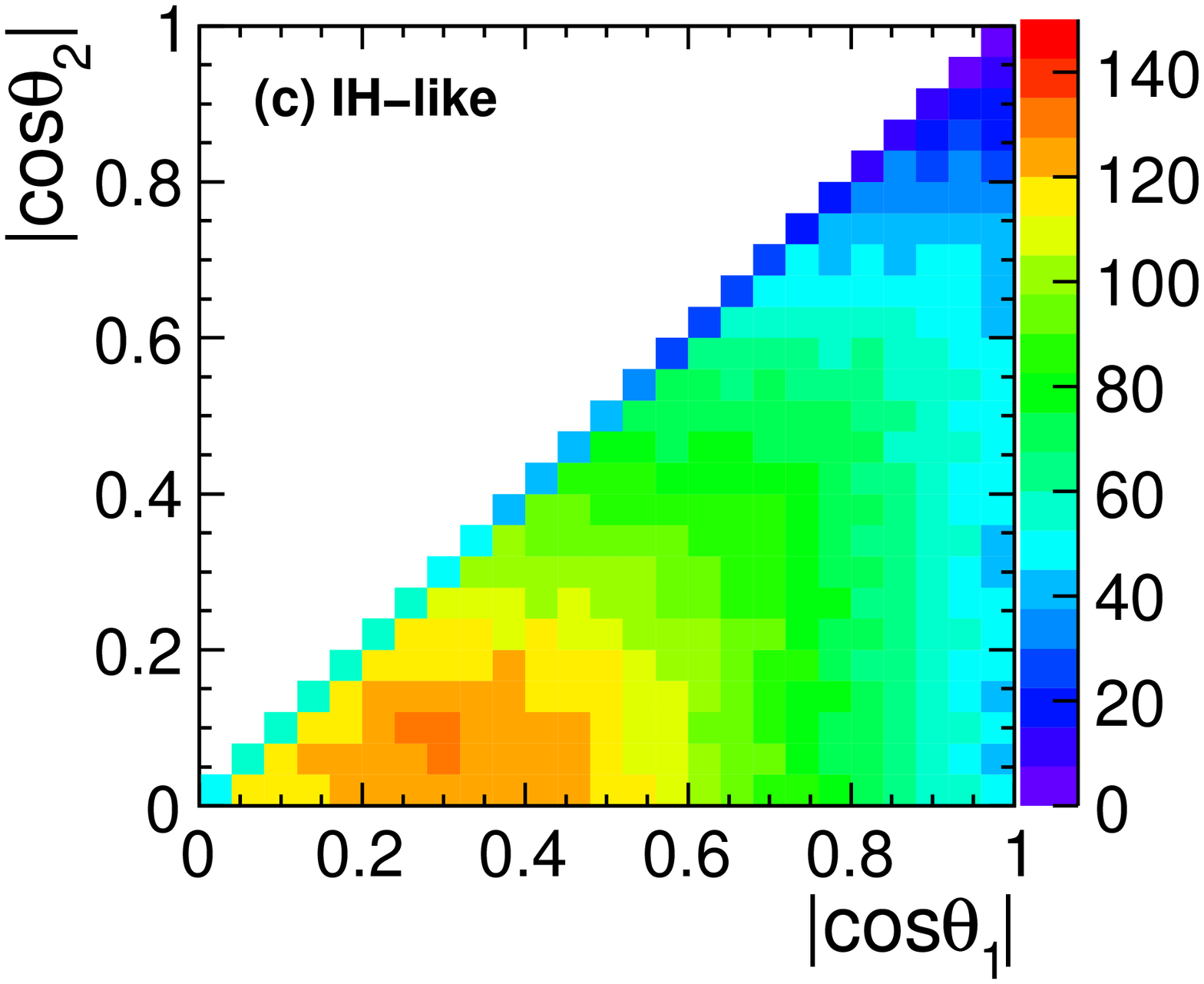}
\end{center}
\end{minipage}
\begin{minipage}{.45\textwidth}
\begin{center}
\includegraphics[width=1\textwidth]{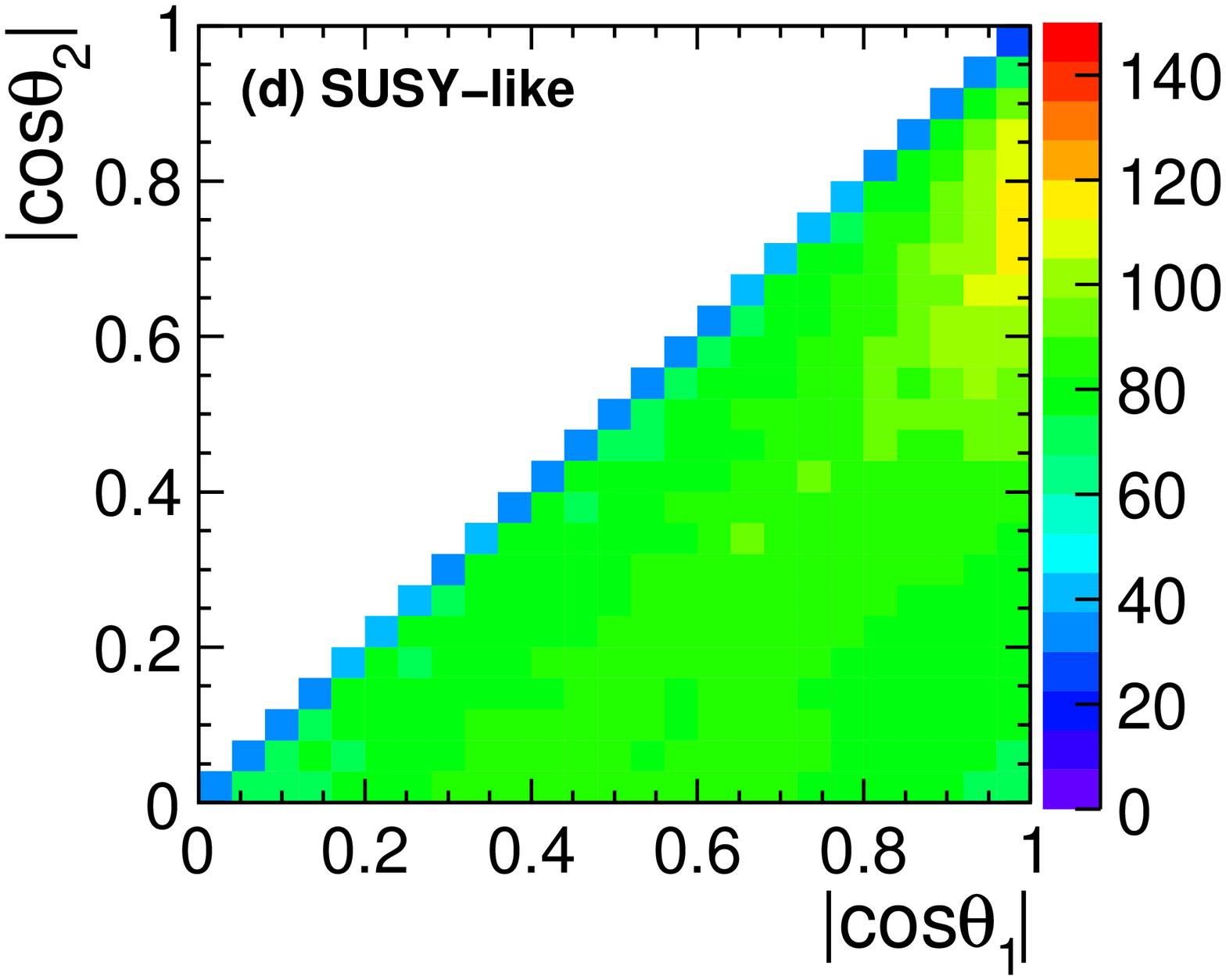}
\end{center}
\end{minipage}
\begin{minipage}{.45\textwidth}
\begin{center}
\includegraphics[width=1\textwidth]{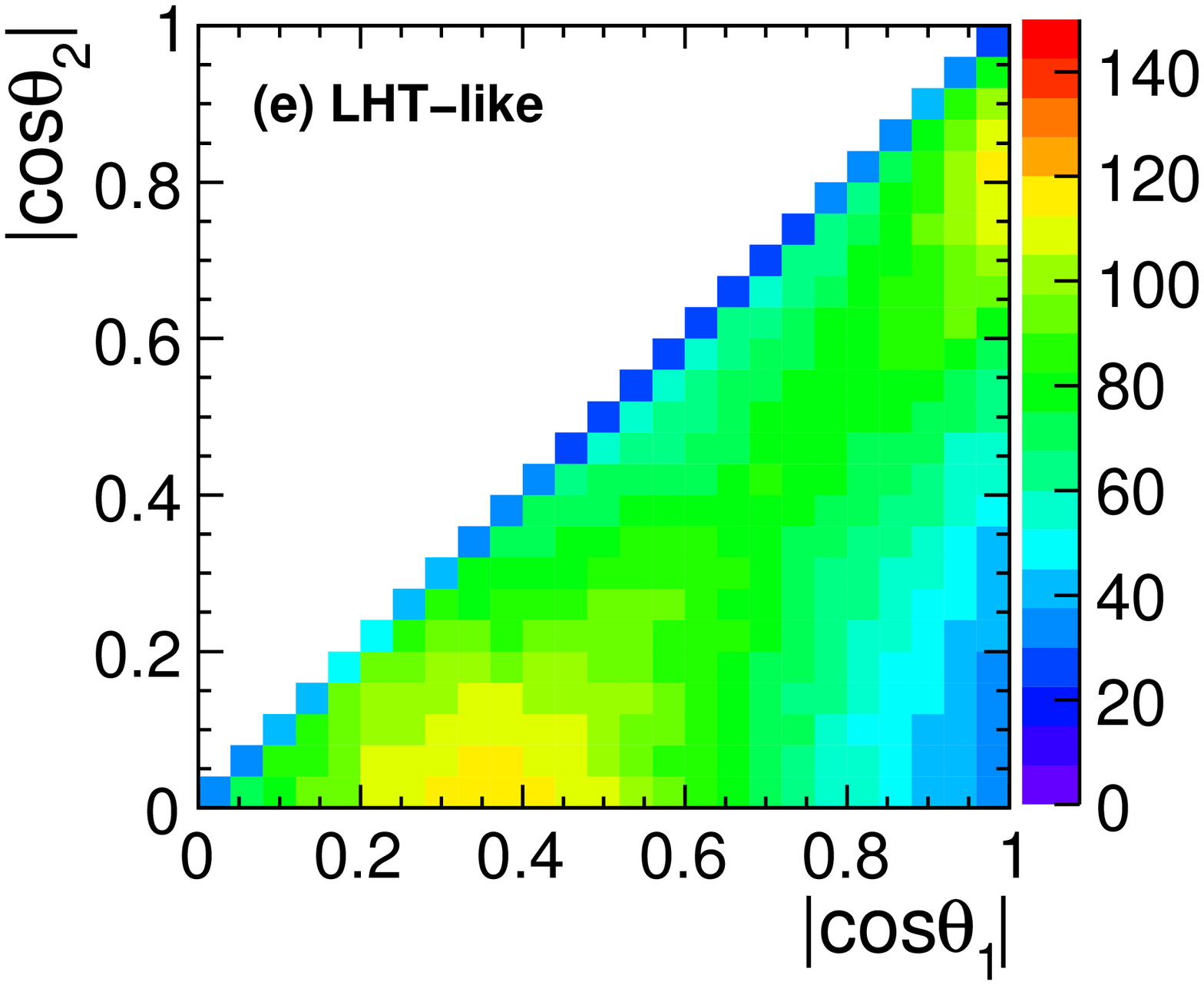}
\end{center}
\end{minipage}
\begin{minipage}{.45\textwidth}
\begin{center}
\includegraphics[width=1\textwidth]{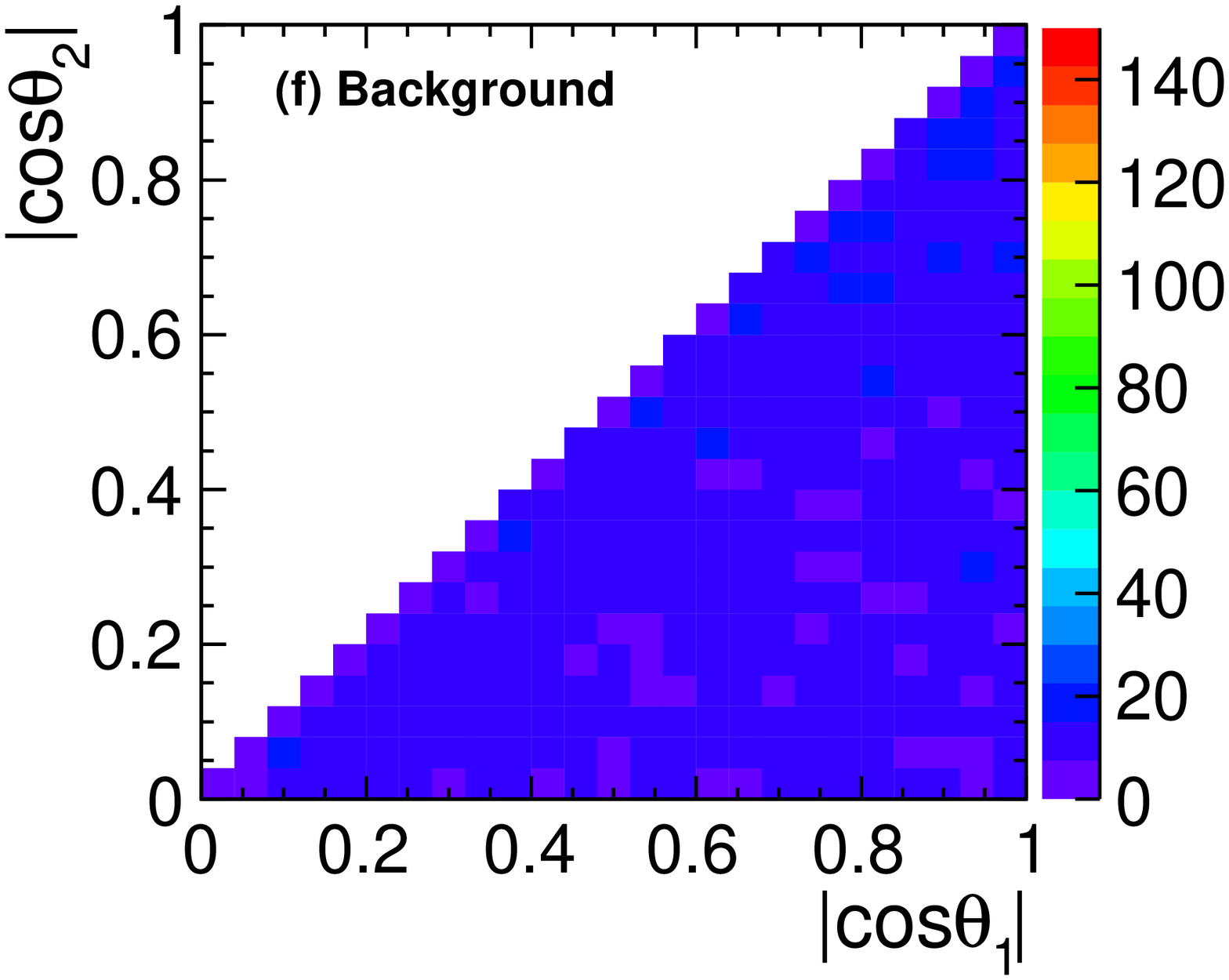}
\end{center}
\end{minipage}
\end{center}
\caption{Production angle distributions in the Point II study.
(a) and (b) show the generated and the reconstructed 1-dimensional distributions. (a) shows the true distribution, while (b) includes both of the two solutions.
(c)-(f) give 2-dimensional distributions of the IH-like, the SUSY-like, the LHT-like, and the SM background, normalized to $\sigma_s = 200$ fb and $\mathcal{L}_\mathrm{int} = 500$ fb$^{-1}$.
}
\label{fig:angle1000}
\end{figure}

\begin{figure}
\begin{center}

\begin{minipage}{.45\textwidth}
\begin{center}
\includegraphics[width=1\textwidth ]{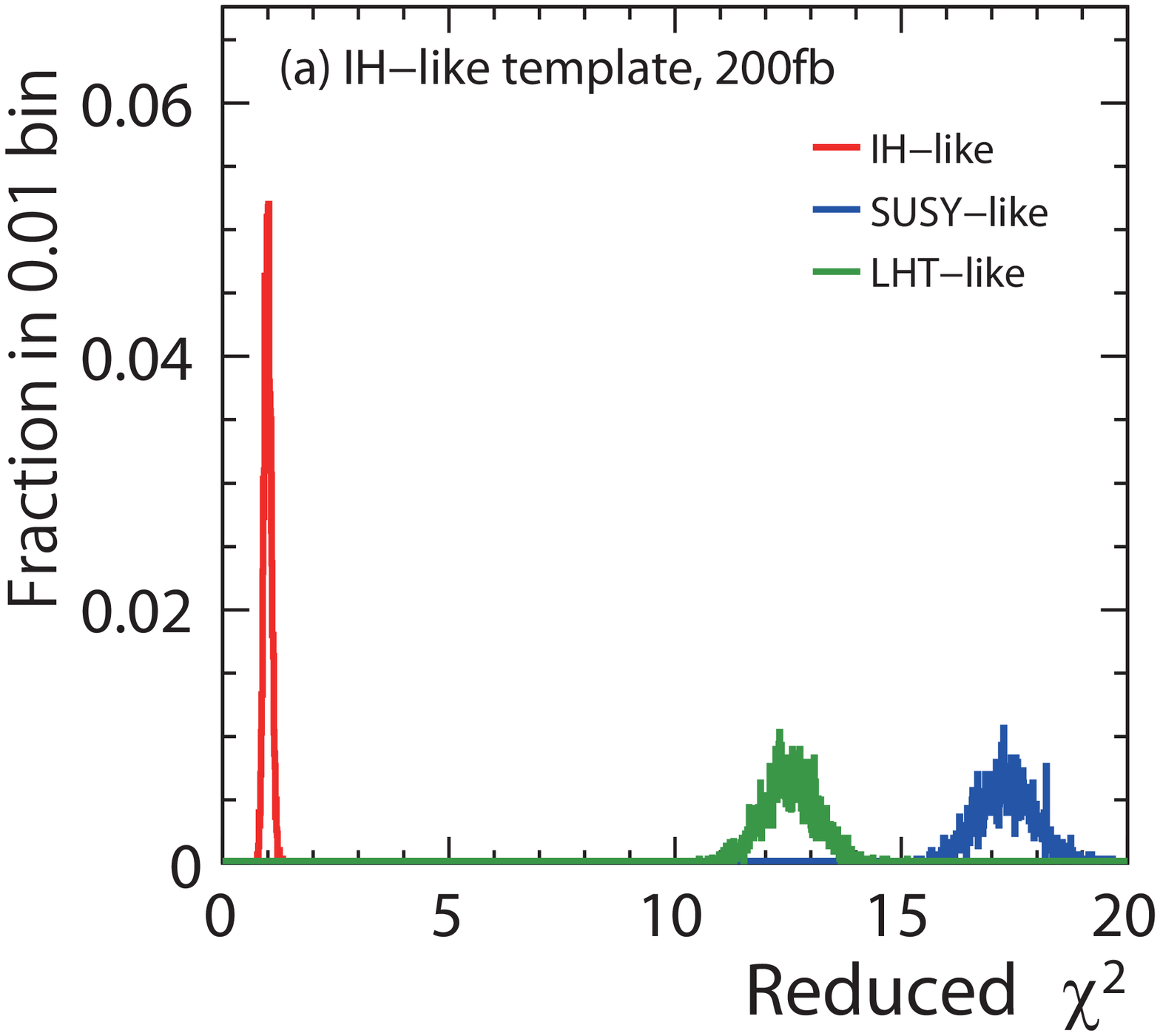}
\end{center}
\end{minipage}
\begin{minipage}{.45\textwidth}
\begin{center}
\includegraphics[width=1\textwidth ]{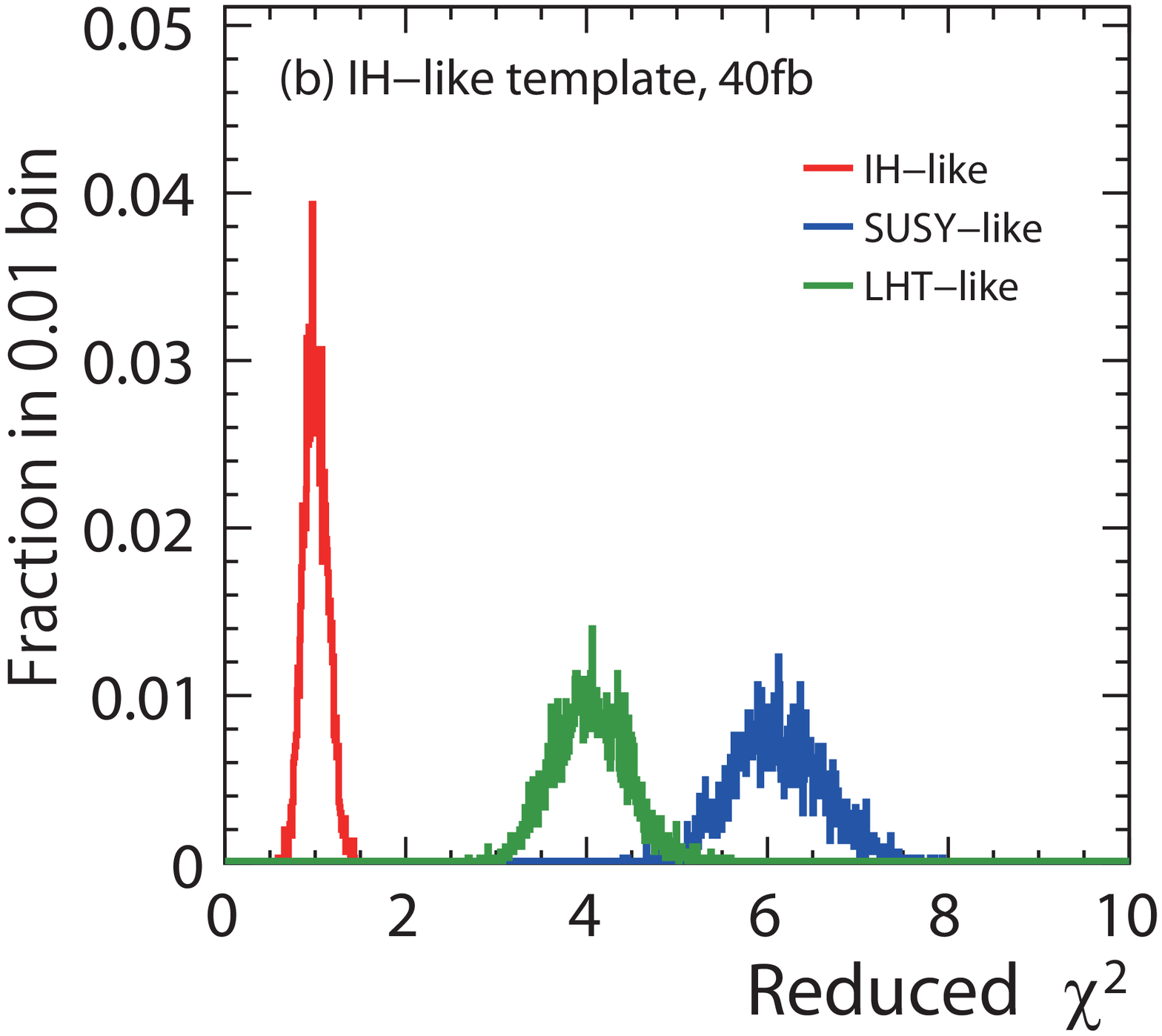}
\end{center}
\end{minipage}
\begin{minipage}{.45\textwidth}
\begin{center}
\includegraphics[width=1\textwidth ]{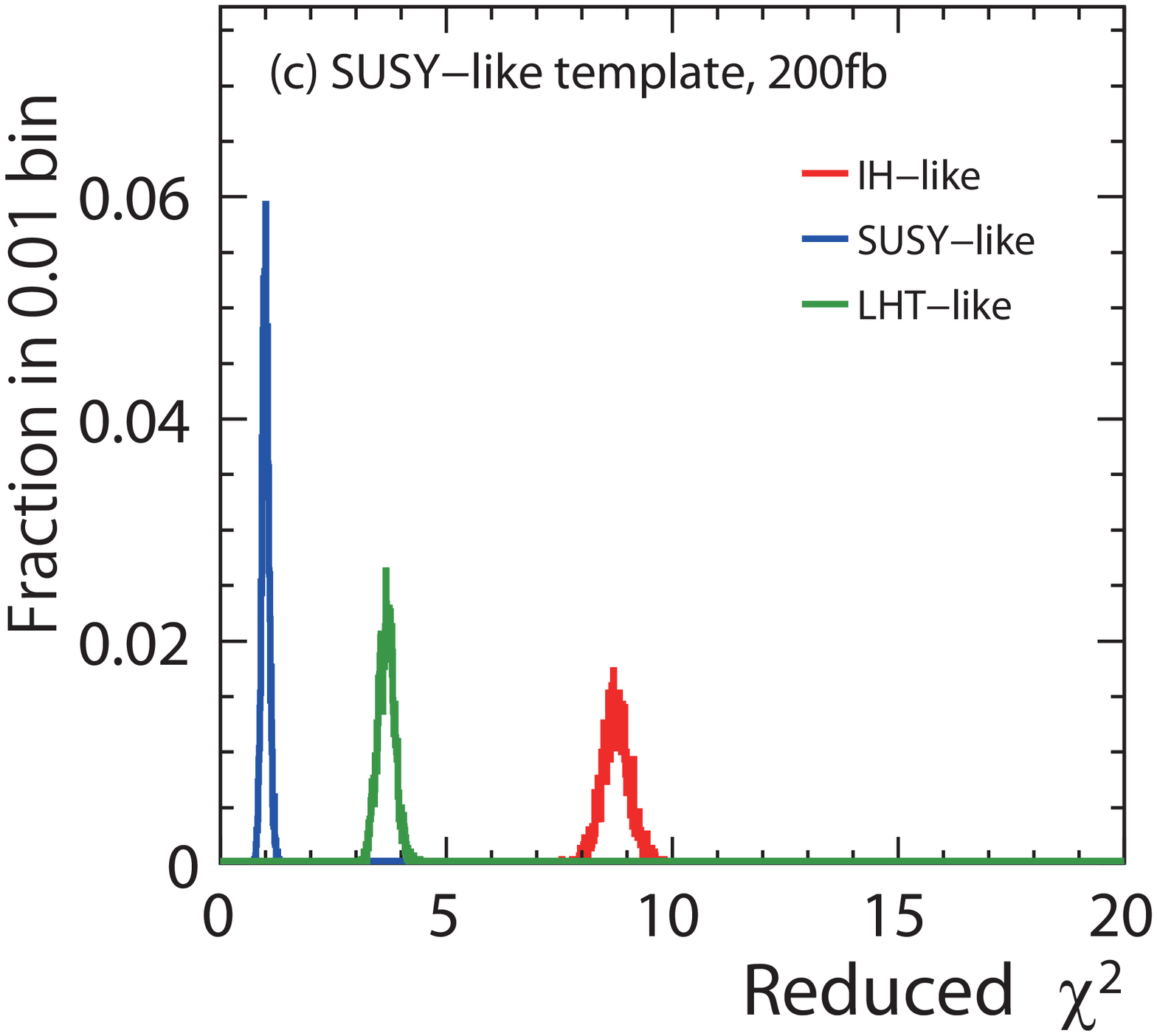}
\end{center}
\end{minipage}
\begin{minipage}{.45\textwidth}
\begin{center}
\includegraphics[width=1\textwidth ]{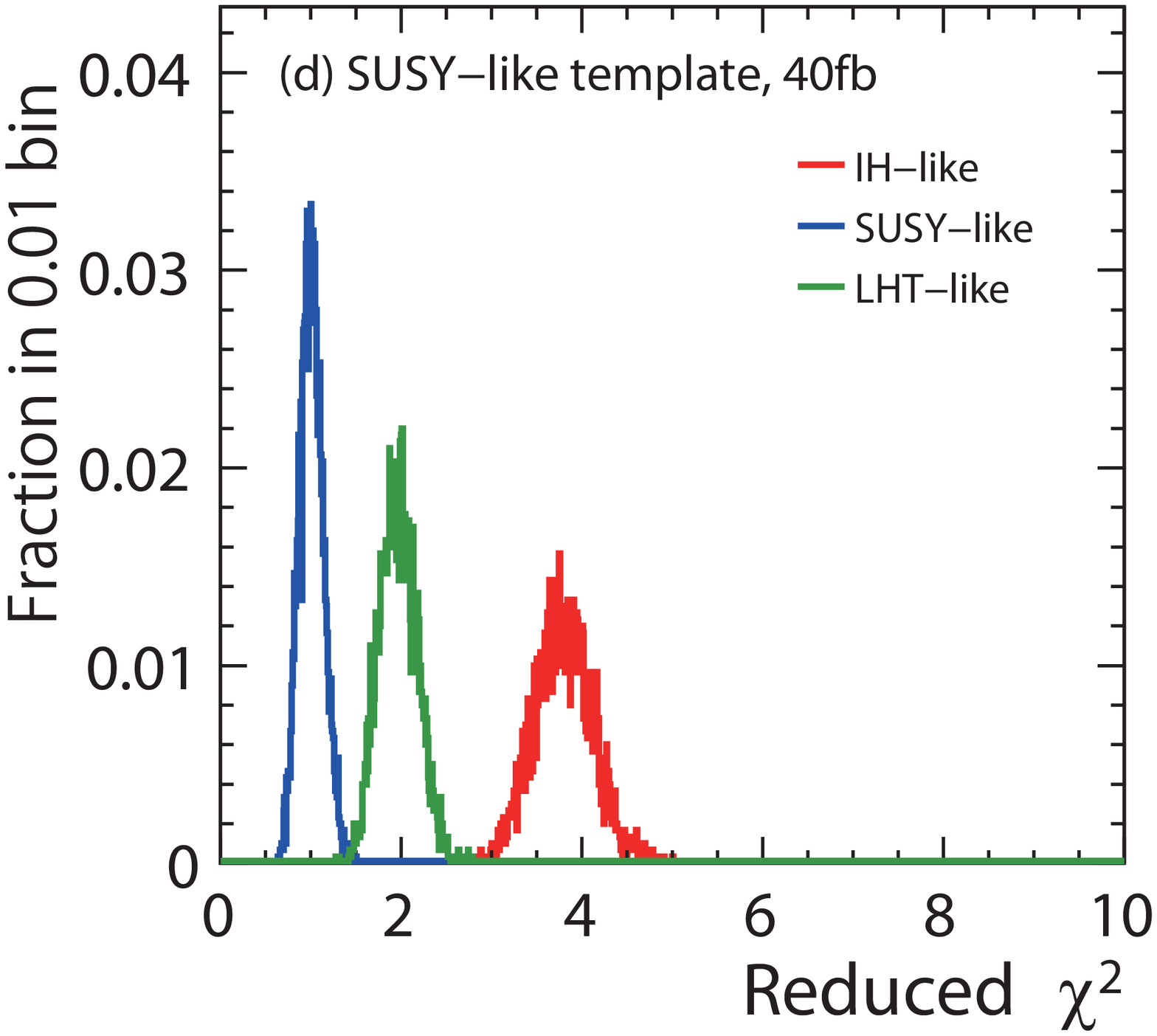}
\end{center}
\end{minipage}
\begin{minipage}{.45\textwidth}
\begin{center}
\includegraphics[width=1\textwidth ]{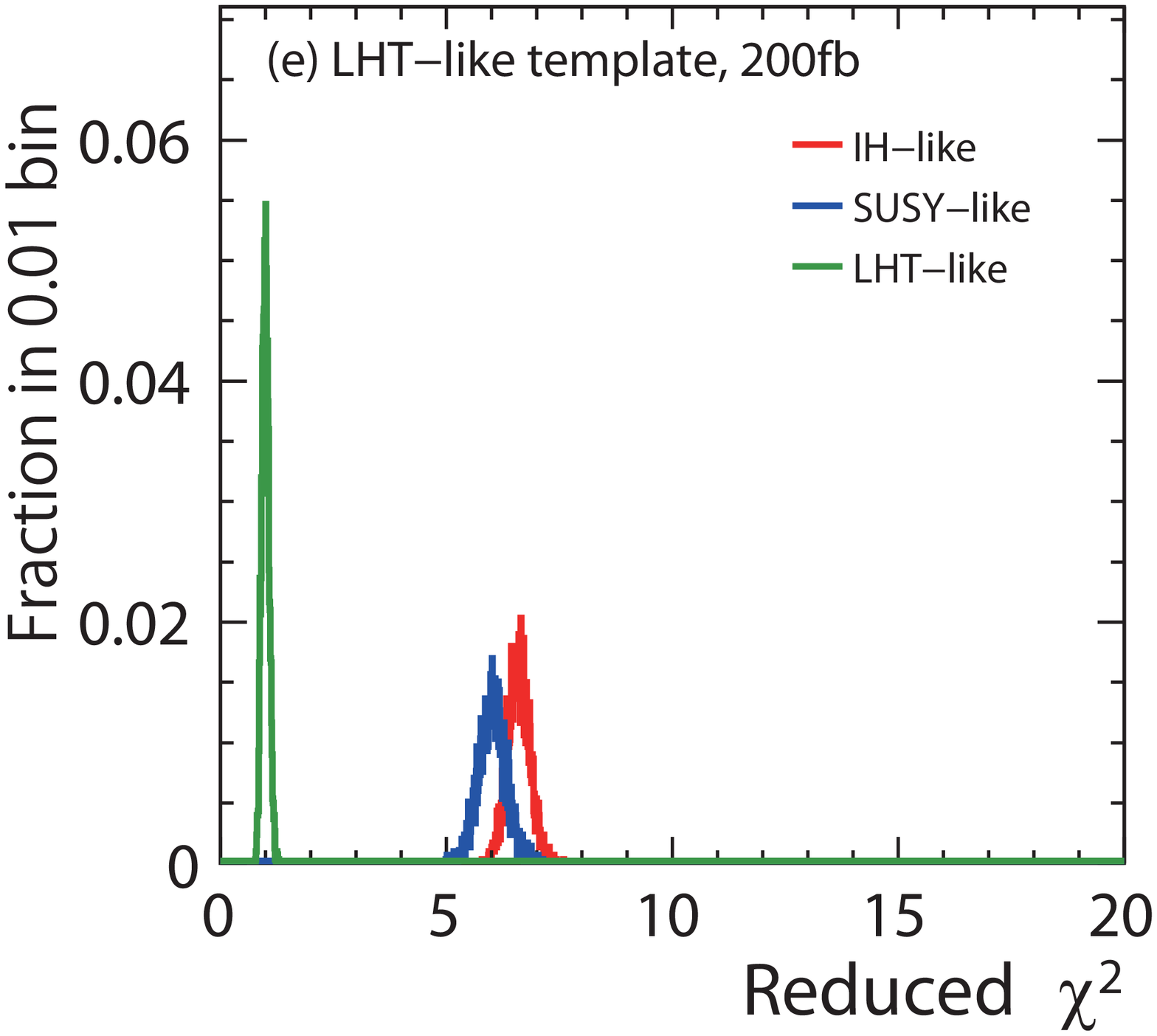}
{\small (a) 200 fb}
\end{center}
\end{minipage}
\begin{minipage}{.45\textwidth}
\begin{center}
\includegraphics[width=1\textwidth ]{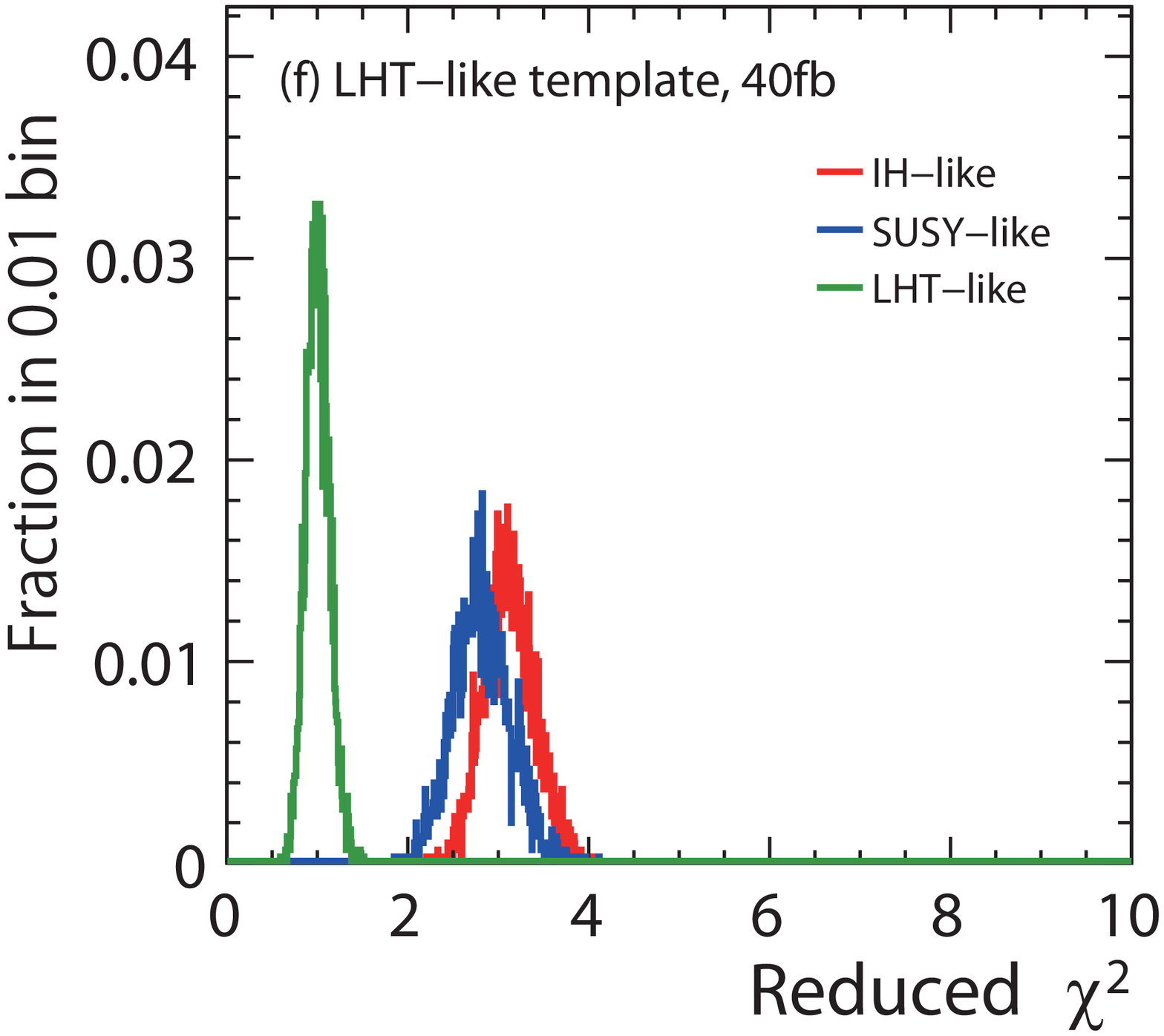}
{\small (b) 40 fb}
\end{center}
\end{minipage}
\caption{$\tilde{\chi^2}(M_D, M_T)$ distributions with $\mathcal{L}_\mathrm{int} = 500$ fb$^{-1}$ in the Point II study.}	
\label{fig:chi2dist1000}
\end{center}
\end{figure}

The model separation was studied using the two-dimensional production angle distributions as in the Point I study.
Figure \ref{fig:angle1000} shows the one- and two-dimensional histograms for the two solutions of the production angle.
The angular distributions for each physics model were prepared with a high-statistics sample, and normalized to
$\mathcal{L}_\mathrm{int} = 500$ fb$^{-1}$.
The number of bins in 2-dimensional histograms ($N$ in Eq.~(\ref{eq:chi2}) ) is 325 (instead of 210 in the Point I study).
Figure \ref{fig:chi2dist1000} shows the $\tilde{\chi^2}$ distributions and Table \ref{tb:achi2_1tev} tabulates the expectation values of separation power $\bar{P}$ for each physics model, which are defined in Eq.~(\ref{eq:chi2}).
The physics model can be identified confidently by using the $\bar{P}$ values in both of the $\sigma_s = 200$ fb and 40 fb cases.

\begin{table}
\center{
\begin{tabular}{|l|r|r|r|r|}
\hline
$\sigma_s$  & $M_D \backslash M_T$ & IH-like & SUSY-like & LHT-like \\
\hline
200 fb &  IH-like  &  -  & 109 & 82.5 \\
       & SUSY-like & 216 & -   & 75.0 \\
       & LHT-like  & 156 & 46.0  & - \\  \hline 
40 fb   &  IH-like& -    & 29.1 & 23.9 \\
       & SUSY-like & 46.8 & -    & 21.6 \\
       & LHT-like  & 30.9 & 9.38 & - \\
\hline
\end{tabular}
}

\caption{Expectation values of separation power $\bar{P}$ between three models with the 2-dimensional production angle distribution with $\mathcal{L}_\mathrm{int} = 500$ fb$^{-1}$ in the Point II study.}
 \label{tb:achi2_1tev}
\end{table}

\subsubsection{Threshold scan}
\begin{figure}
\begin{center}
\begin{minipage}{.32\textwidth}
\begin{center}
\includegraphics[width=1\textwidth]{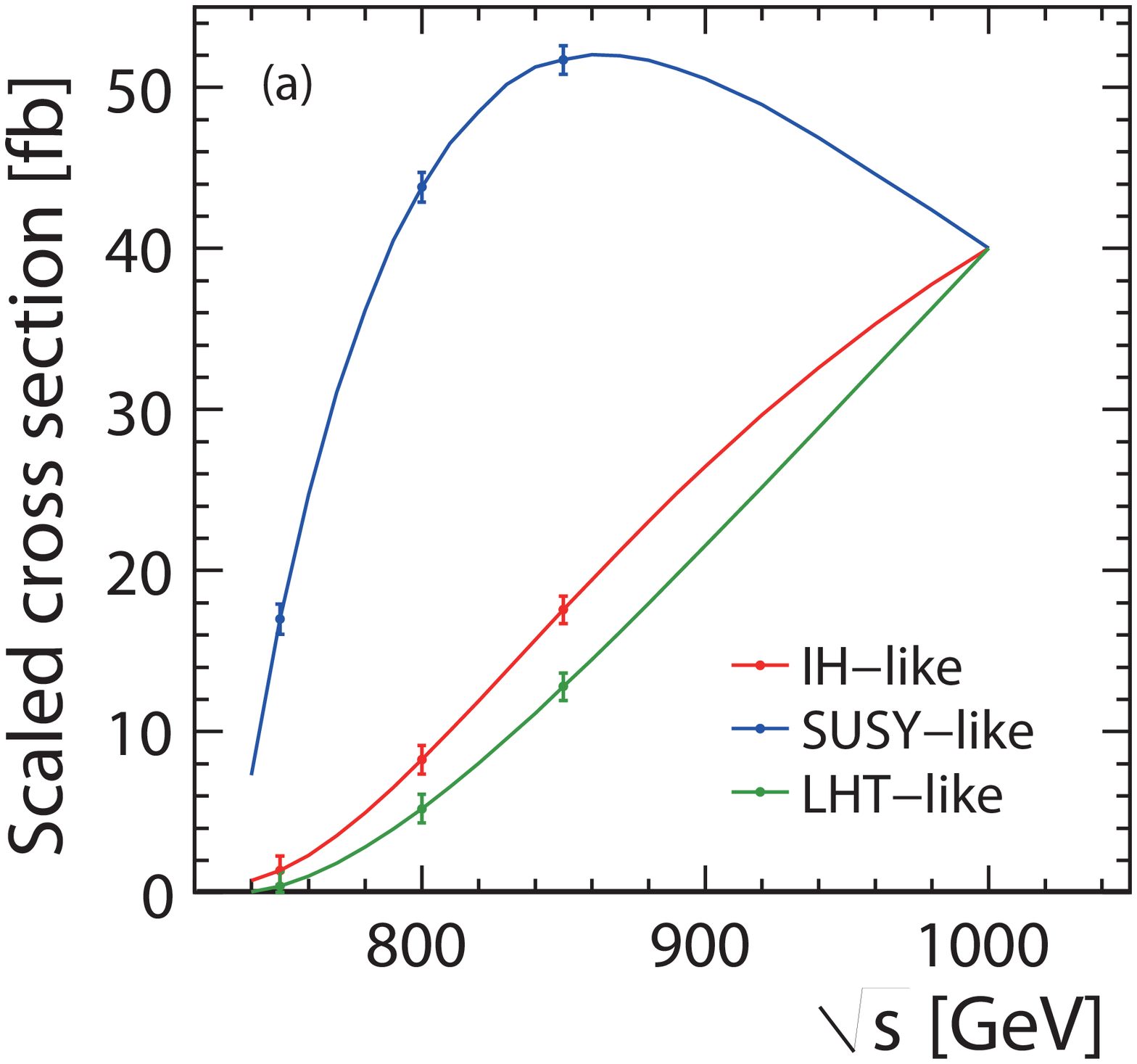}
\end{center}
\end{minipage}
\begin{minipage}{.32\textwidth}
\begin{center}
\includegraphics[width=1\textwidth]{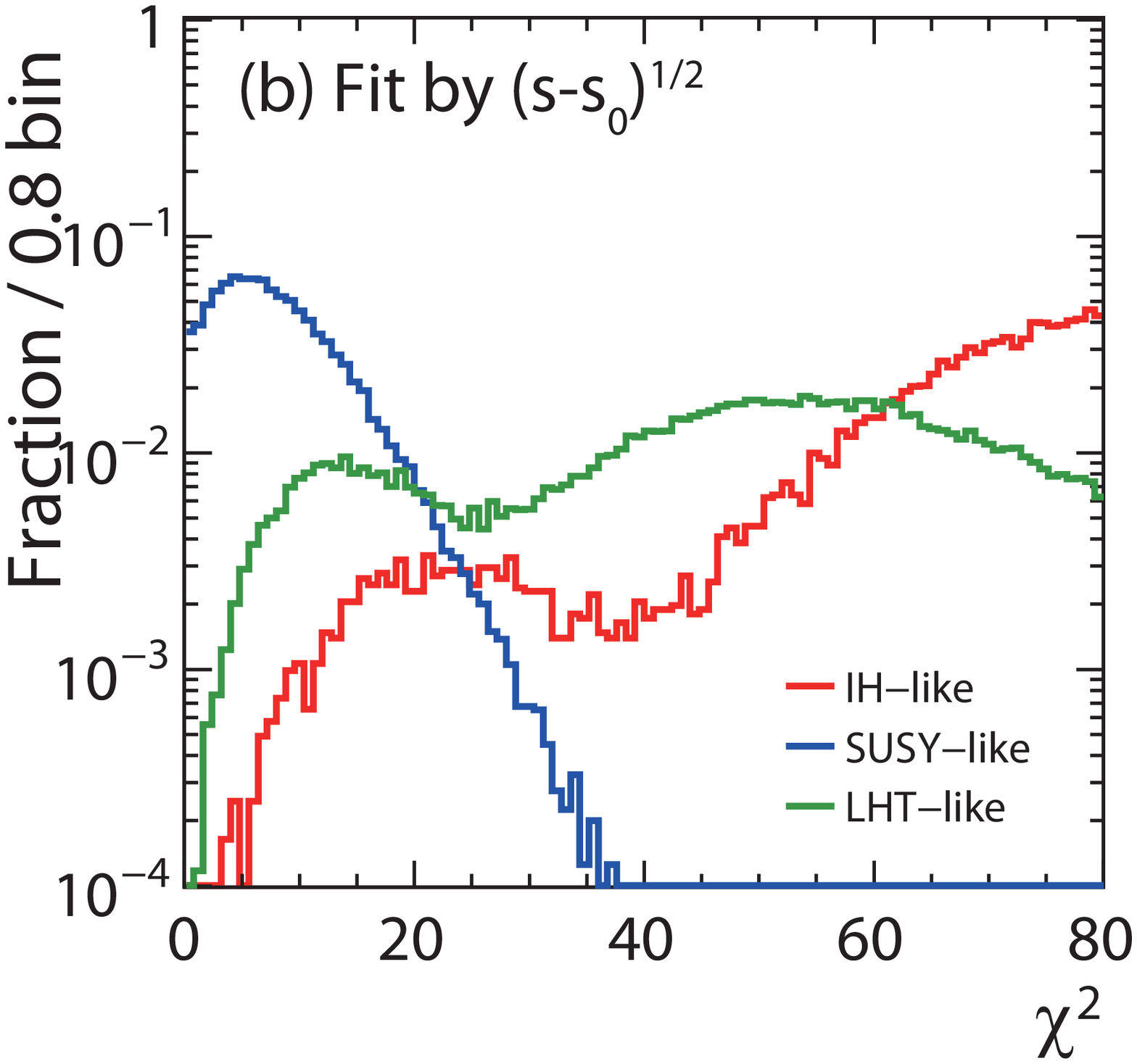}
\end{center}
\end{minipage}
\begin{minipage}{.32\textwidth}
\begin{center} \includegraphics[width=1\textwidth]{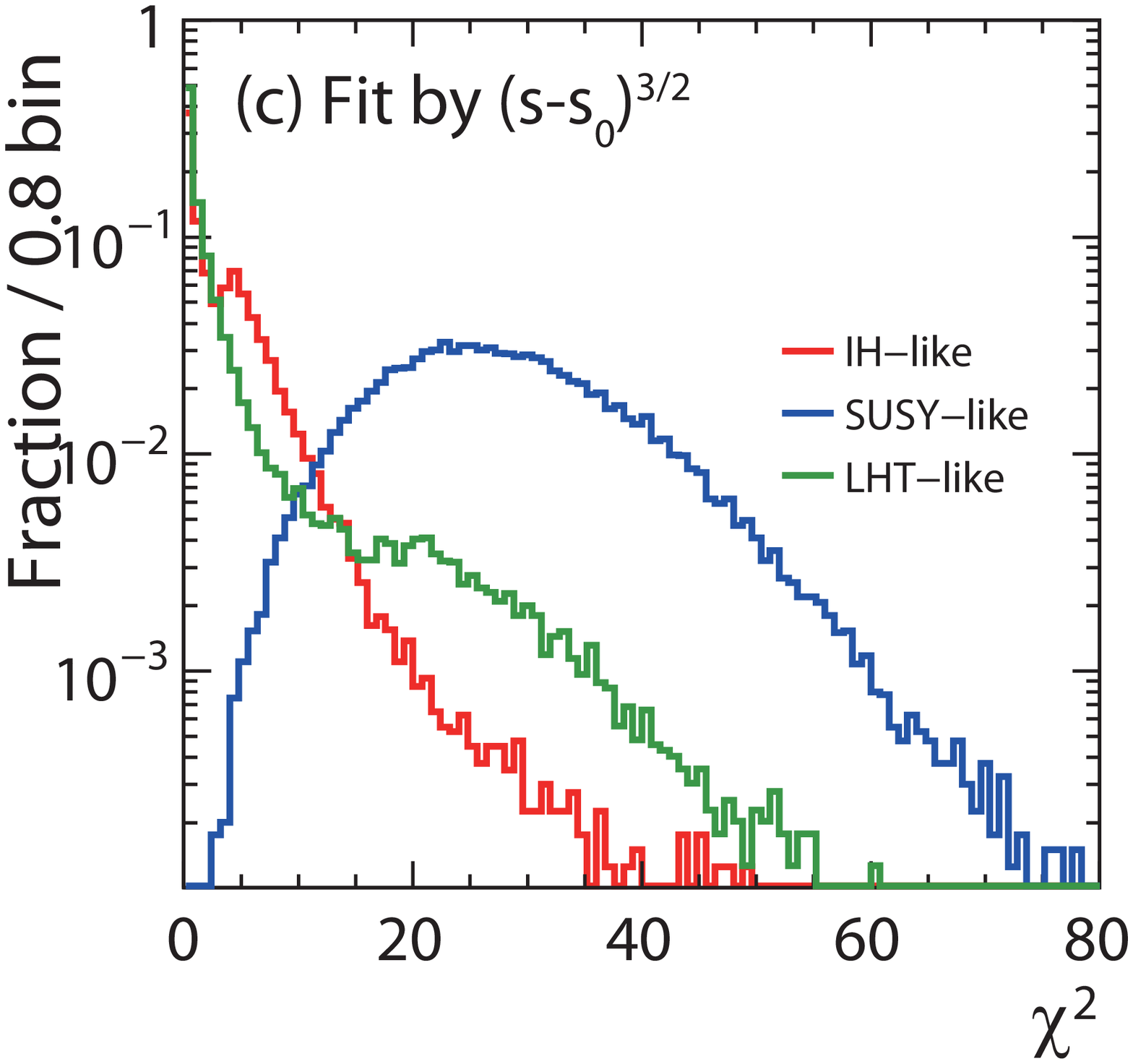}
\end{center}
\end{minipage}
\end{center}
\caption{(a) Dependence of the cross section on the center of mass energy, normalized to $\sigma_s = 40$ fb at 1 TeV in the Point II study. Error bars are given assuming $\mathcal{L}_\mathrm{int} = 50$ fb$^{-1}$ data at each point.  (b),(c) Result of the $\chi^2$ fit for the $(s - s_0)^{1/2}$ case and the $(s - s_0)^{3/2}$ case, respectively.
}
\label{fig:thscan1000}
\end{figure}

Threshold scan was also performed with the same procedure as in the Point I study.
Figure \ref{fig:thscan1000} (a) shows the $\sqrt{s}$ dependence of the cross section of each model.
We performed a 3-point toy-MC scan of $\sqrt{s} = $ 750, 800, and 850 GeV with $\mathcal{L}_\mathrm{int} = 50$ fb$^{-1}$ at each point.
The signal cross section $\sigma_s$ was scaled to 40 fb at $\sqrt{s}$ = 1 TeV.
The cut efficiency and background cross section are assumed to be the same as those at $\sqrt{s}$ = 1 TeV.
The $\chi^2$ distributions of fits to Eq.~(\ref{eq:threshold}) with $n = 1/2$ and $3/2$, shown in Figs.~\ref{fig:thscan1000} (b),(c),
were obtained by the same methods as in the Point I study.
Both distributions give good separation between the SUSY-like and the other two models.
If we assume events with $\chi^{2} < 16$ as the SUSY-like events for the fitting with the power of $1/2$, the IH-like and the LHT-like events were disfavored with probability of 99.4\% and 90.7\%, respectively. Here, 92.0\% of the SUSY events were selected as the SUSY-like events. On the other hand, if we assume events with $\chi^{2} > 13$ as the SUSY-like events for fit with the power of $3/2$, the IH-like and the LHT-like events were disfavored with probability of 97.5\% and 90.8\%, respectively, and 92.3\% of the SUSY-like events were selected as the SUSY-like events. 

%% file: Summary.tex
\section{Summary}
\label{sec:summary}

The WIMP dark matter is one of important candidates predicted in many new physics models at the TeV scale, which will be detected at the ILC. Interestingly, various new physics models predict the existence of the process $e^+ e^- \to \chi^+ \chi^- \to W^+ W^- \chi ^0 \chi^0$, which allows us to measure properties of the dark matter ($\chi^0$) and the new charged particle ($\chi^\pm$) with good accuracy. With the use of the process, it is also possible to discriminate the new physics models in a model-independent way. We have shown that the masses of $\chi^0$ and  $\chi^\pm$, the angular distribution of $\chi^\pm$, and the threshold behavior of the $\chi^\pm$ production cross section can be accurately measured at the ILC. In fact, it was shown quantitatively that these measurements can be used to discriminate the new physics models: IH-like, SUSY-like, and LHT-like models.

In the study of the benchmark point I, it turns out that the masses of $\chi^0$ and $\chi^\pm$ are determined with accuracies of 5\% and 0.2\% when the production cross section of $\chi^\pm$ is $\sigma = 40$ fb, and 2\% and 0.04\% when $\sigma = 200$ fb. The measurement of the angular distribution of $\chi^\pm$ enables us to discriminate the IH-like model from the other models, while the SUSY-like model can be discriminated by the threshold scan of the process when $\sigma = 40$ fb. When $\sigma = 200$ fb, all the new physics models can be separated from each other by using only the measurement of the angular distributions. On the other hand, in the study of the benchmark point II, the masses of $\chi^0$ and $\chi^\pm$ are determined with accuracies of 5\% and 0.8\% when the production cross section of $\chi^\pm$ is $\sigma = 40$ fb, and 2\% and 0.2\% when $\sigma = 200$ fb. The new physics models can be discriminated by using the angular distribution even if $\sigma = 40$ pb.

In this article, we have shown that new physics models (IH-, SUSY-, and LHT-like models) can be discriminated at the ILC. On the other hand, it is also true that we need to extend the method developed in this article in order to establish a strategy for the discrimination in a completely model-independent way. For example, the angular distribution of the $\chi^+\chi^-$ production would be changed if there is a diagram in which a new particle (such as selectron in MSSM or heavy electron in LHT models) propagates in $t$-channel. Even if the mass of such a new particle is as heavy as 1 TeV, its effect can be sizable in general and the resultant production angle distribution may become significantly asymmetric. In addition, if we allow a more generic (Lorentz) structure for the $\chi^+\chi^-Z$ vertex, the angler distribution may also be affected. In these situations, the identification of the $W$ charge [24] becomes very important to reconstruct the asymmetric distribution. Moreover, the beam polarization and the measurements of the $W$ polarization and the $W$ energy distribution may also play an essential role to extract the information on the vertices involving new particles; these will help us not only to discriminate new physics models but also to determine the properties of the WIMP dark matter in detail.